\documentclass[10pt,letterpaper]{article}
\usepackage[top=0.85in,left=2.75in,footskip=0.75in]{geometry}

\usepackage{amsmath,amssymb}

\usepackage{changepage}

\usepackage{textcomp,marvosym}

\usepackage{cite}

\usepackage{nameref,hyperref}

\usepackage[nopatch=eqnum]{microtype}
\DisableLigatures[f]{encoding = *, family = * }

\usepackage[table]{xcolor}

\usepackage{array}

\newcolumntype{+}{!{\vrule width 2pt}}

\newlength\savedwidth

\raggedright
\usepackage[aboveskip=1pt,labelfont=bf,labelsep=period,justification=raggedright,singlelinecheck=off]{caption}

\makeatletter
\renewcommand{\@biblabel}[1]{\quad#1.}
\makeatother

\usepackage{lastpage,fancyhdr,graphicx}
\usepackage{epstopdf}
\fancyheadoffset[L]{2.25in}
\fancyfootoffset[L]{2.25in}
\begin{document}
\vspace*{0.2in}

\begin{flushleft}
{\Large
\textbf{A beam--membrane biomechanical vocal fold model incorporating posturing and glottal conformation} 
}
\newline
\\
Mohamed A. Serry\textsuperscript{1,2},
Mat\'ias Za\~nartu\textsuperscript{3},
Sean D. Peterson\textsuperscript{1*}
\\
\bigskip
\textbf{1} Department of Mechanical and Mechatronics Engineering, University of Waterloo, Waterloo, Ontario N2L 3G1, Canada
\\
\textbf{2} Department of Applied Mathematics, University of Waterloo, Waterloo, Ontario N2L 3G1, Canada\\
\textbf{3} Department of Electronic Engineering, Universidad T\'ecnica Federico Santa Mar\'ia, Valpara\'iso, Chile
\\
\bigskip

* peterson@uwaterloo.ca

\end{flushleft}

\section*{Abstract}
The posture of the vocal folds produced by laryngeal muscle activation plays a central role in determining the dynamics of voice production. Abnormal vocal fold configurations are frequently associated with inefficient phonation and a variety of voice disorders. Although diverse glottal closure patterns have been observed clinically, the biomechanical mechanisms governing their dynamic behavior and resulting phonatory characteristics remain incompletely understood. Moreover, existing numerical models that incorporate the effects of the intrinsic musculature on posturing and glottal conformation are computationally expensive, which limits their suitability for large-scale parametric investigations. In this work, we introduce a computationally inexpensive vocal fold (VF) model wherein the body and cover VF layers are treated as a composite beam and a coupled membrane, respectively. Intrinsic laryngeal muscle activation, in addition to positioning the arytenoid cartilages and cricothyroid joint, introduces moments at the boundaries of the structure that influence glottal conformation. The model produces phonatory characteristics that are qualitatively consistent with those reported in high-fidelity finite-element models and clinical studies, thereby supporting its predictive capability while offering substantial  computational advantage. The proposed framework provides biomechanical insights into the influence of incomplete glottal closure on phonation dynamics and may serve as a computationally tractable tool for investigating mechanisms underlying certain voice disorders.

\section*{Author summary}
 The human voice is produced when the vocal folds vibrate as air flows through the glottis (the area between the folds). Studies have shown that the shape of the glottis strongly influences vocal fold vibration, internal stresses, and contact pressure, which are metrics influencing vocal fold health. The intrinsic laryngeal muscles position and stretch the vocal folds, impacting the static and dynamic glottal geometries. Although high-fidelity finite element computer models can be used to study these processes, they require substantial computational resources, making them ill-suited to explore broad physiological conditions. In this work, we develop a computationally efficient biomechanical model that captures the influence of laryngeal muscle activation on vocal fold motion, glottal conformation, and voice production. The model reproduces many trends reported in experimental studies and high-fidelity simulations while incurring substantially less computational expense. By providing a practical tool for exploring the biomechanics of phonation, this framework may help improve our understanding of both normal voice production and voice disorders.
\clearpage
\newgeometry{top=0.85in,left=1in,right=1in,footskip=0.75in}

\section*{Introduction}
Vocal fold (VF) posturing is a key determinant of voice production characteristics. VF configuration and geometry are governed by the specific combination and relative activation of the intrinsic and extrinsic laryngeal muscles, which generate internal forces and bending moments that shape the folds \cite{ChhetriNeubauerBerry12, serry2023modeling}. Aerodynamic loading and VF collision during phonation further influence VF dynamics. Certain glottal configurations may lead to inefficient phonation \cite{ZanartuGalindoErathPetersonWodickaHillman14} and, in some cases, elevated stress concentrations within the VFs that may contribute to tissue trauma \cite{DejonckereKob09}. Accordingly, VF posturing and geometry play an essential role in understanding voice disorders associated with inefficient vocal function or phonotraumatic damage, including Parkinson’s disease \cite{HansonGerrattWard84}, muscle tension dysphonia \cite{MorrisonRammage93}, and non-phonotraumatic or phonotraumatic vocal hyperfunction \cite{HillmanSteppVanStanZanartuMehta20}. 

The VFs exhibit a wide variety of static (\textit{i.e.}, prephonatory) and dynamic (phonatory) configurations, including concave shapes with a posterior glottal opening, bowed shapes with an anterior opening, hourglass-shaped configurations, and spindle-shaped glottal patterns \cite{RajaeiBarzegarMojiriNilforoush14,SoderstenHertegardHammarberg95,MorrisonRammage93,NguyenKennyTranLivesey09}. Experimental and clinical studies have provided insight into the laryngeal mechanisms responsible for such curved VF geometries, demonstrating that these configurations, particularly in static settings, arise, at least in part, from laryngeal maneuvers involving different combinations of laryngeal muscle activation \cite{ChhetriNeubauerBerry12}.  For example, inspection of cadaver larynges has suggested that posterior glottal opening can be induced  by activation of the posterior cricoarytenoid (PCA) muscle \cite{MorrisonRammage93}. Experimental investigations using excised canine larynges further demonstrated that activation of the thyroarytenoid (TA) muscle while maintaining low activation of the primary adductory muscles, namely the lateral cricoarytenoid (LCA) and interarytenoid (IA) muscles, can produce posterior glottal opening through combined anterior and mid-membranous closure \cite{ChoiYeBerkeKreiman93}. Conversely, a relaxed TA muscle combined with sufficient activation of the adductory muscles can produce an anterior glottal opening and bowed VF configurations \cite{ChoiYeBerkeKreiman93,ChhetriNeubauer15}. Full membranous closure generally requires co-activation of the TA and adductory muscles \cite{ChhetriNeubauer15}. Furthermore, canine studies demonstrated that, for fixed TA activation, increasing activation of the cricothyroid (CT) muscle tends to straighten the folds, thereby reducing medial bulging induced by TA activation and mitigating posterior constriction associated with LCA and IA activation, ultimately increasing the glottal area \cite{ChhetriNeubauerBerry12}. Interestingly, certain patterns of co-activation involving the LCA, IA, and TA muscles were also found to produce static hourglass-shaped glottal configurations \cite{ChhetriNeubauerBerry12}.

Several numerical investigations have provided further insights into curved and incomplete glottal closure patterns. For example, high-fidelity computational simulations have been used to study the influence of laryngeal muscle activation on resting glottal geometry, yielding results that are consistent with experimental observations regarding the roles of the LCA, IA, and TA muscles in shaping the glottis \cite{YinZhang14,YinZhang16}. Interestingly, it was later demonstrated that many of these muscle-induced geometric effects can be qualitatively predicted using a simplified static beam model, in which laryngeal muscle activation generates internal bending moments that shape the VFs \cite{serry2023euler}. More recent studies have employed physiologically detailed finite-element biomechanical VF models incorporating laryngeal muscle activation \cite{MovahhediGengXueZheng21,JiangGengZhengXue24}. Although these studies did not explicitly focus on glottal geometry, they provide valuable insights into how synergistic activation of laryngeal muscles influences phonatory measures and the mechanical behavior of the VFs.

Despite these important contributions, a comprehensive understanding of how laryngeal muscle activation influences dynamic phonatory behavior while accounting for VF conformation remains incomplete. Experimental studies often face challenges in isolating contributing biomechanical factors and obtaining consistent phonatory measurements. High-fidelity computational models, although capable of capturing detailed fluid--structure interactions and tissue mechanics, frequently lack mechanical transparency and typically involve substantial computational cost, thereby limiting their suitability for large-scale parametric investigations. It should also be noted that several reduced-order phonation modeling frameworks have sought to relate laryngeal muscle activation to key glottal geometric characteristics, including the glottal gap, VF length and width, and glottal convergence angle \cite{TitzeStory02,AlzamendiPetersonErathHillmanZanartu21}. However, these models suffer from two main limitations. First, the relationship between muscle activation and the resulting lumped mechanical properties is often introduced through heuristic or \textit{ad hoc} rules (see, \textit{e.g.}, \cite{TitzeStory02}), making it difficult to identify, interpret, and validate the underlying biomechanical mechanisms. Second, such models typically assume highly simplified glottal geometries, such as rectangular \cite{TitzeStory02} or triangular \cite{AlzamendiPetersonErathHillmanZanartu21} configurations, and are therefore unable to predict physiologically realistic VF shapes, glottal configurations, and closure patterns.

Motivated by these limitations,  we propose an efficient beam--membrane phonation model for investigating the influence of glottal configuration on voice production. The proposed framework, built upon our previous static beam VF model \cite{serry2023euler}, enables computationally inexpensive simulations while maintaining biomechanical interpretability. In particular, the model integrates a reduced-order muscle-controlled posturing framework \cite{TitzeHunter07,AlzamendiPetersonErathHillmanZanartu21} to extract physiologically relevant posturing variables, including glottal angle and VF prestrain, which are subsequently incorporated into a coupled dynamic beam--membrane VF formulation. The resulting framework substantially extends the static beam model of \cite{serry2023euler} by incorporating VF dynamics, fluid--structure interaction effects, and phonatory behavior.

In Model Overview, we present the proposed modeling framework, including its principal inputs and outputs. In Results and Discussion, we demonstrate the predictive capability of the proposed framework through comparisons with high-fidelity computational and experimental studies, and present additional parametric investigations into the influence of glottal geometry on phonatory characteristics, together with a discussion of the implications and limitations of the proposed approach. Finally, Materials and Methods provides a detailed and rigorous derivation of the structural equations governing the model.

\section*{Model overview}
\begin{figure}[ht!]
    \centering
    \includegraphics[width=0.7\linewidth]{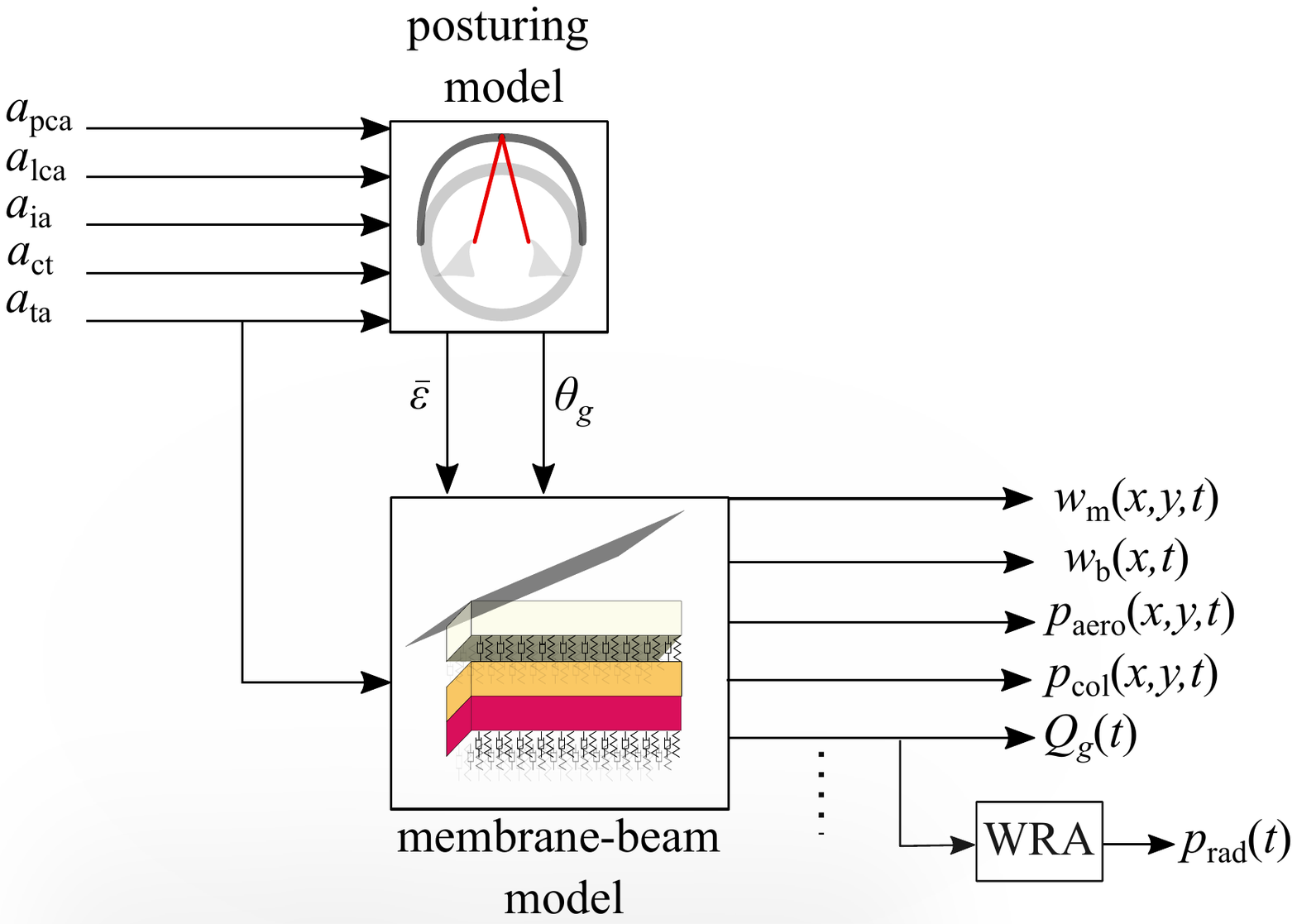}
    \caption{
\textbf{Overview of the proposed phonation modeling framework.}
Schematic representation of the proposed modeling framework, illustrating the flow of information between the posturing model and the beam--membrane phonation model. The posturing model maps laryngeal muscle activation levels to physiologically relevant posturing variables, including vocal fold prestrain and glottal angle. These variables serve as inputs to the beam--membrane model, which predicts vocal fold dynamics and glottal aerodynamics, from which acoustic outputs and phonatory measures are estimated.
}
    \label{fig:ModelOutline}
\end{figure}

The proposed modelling framework aims to relate intrinsic laryngeal muscle activation to VF dynamics during phonation while incorporating the influence of glottal geometry on the resulting vibratory behavior. To this end, the framework assumes initially that muscle activation modifies the positions of the posterior ends of the VFs, which are modeled as line elements, thereby defining a nominally triangular glottal configuration characterized by a prescribed glottal angle $\theta_{g}$ and longitudinal stretch or compression, characterized by nominal strain $\bar{\varepsilon}$. These posturing variables are subsequently incorporated into a dynamic fluid--structure interaction beam--membrane model of the VFs. Posturing parameters induce internal moments within the VFs which, together with aerodynamic and contact forces, govern the resulting VF shapes and vibratory dynamics. An outline of the proposed modelling framework, illustrating the variables transferred from the posturing model to the beam--membrane model together with representative model outputs, is presented in Fig.~\ref{fig:ModelOutline}. 

\subsection*{Posturing model}
The laryngeal posturing framework of \cite{TitzeHunter07} is adopted herein, specifically via the adapted implementation presented by \cite{AlzamendiPetersonErathHillmanZanartu21}\footnote{We adapt the implementation in \cite{AlzamendiPetersonErathHillmanZanartu21} by tuning the constitutive  and lumped parameters to produce physiologically realistic posturing behavior.}. This framework relates intrinsic laryngeal muscle activations to the resulting prephonatory glottal configuration via rotational and translational motions of the cricothyroid joints and arytenoid cartilages.  The VFs and surrounding tissues are represented by spring-like mechanical elements. The posterior ends of the VFs are attached to the arytenoid cartilages at the vocal processes. The resulting cartilage displacements are then used to determine the nominal VF strain $\bar{\varepsilon}$ and the glottal half-angle $\theta_g$.

The inputs to the posturing model are the normalized activation levels of the intrinsic laryngeal muscles,
\[
a_{\mathrm{ta}},\quad
a_{\mathrm{ct}},\quad
a_{\mathrm{lca}},\quad
a_{\mathrm{ia}},\quad
a_{\mathrm{pca}},
\]
corresponding to the thyroarytenoid (TA), cricothyroid (CT), lateral cricoarytenoid (LCA), interarytenoid (IA), and posterior cricoarytenoid (PCA) muscles, respectively.

Although the posturing frameworks of \cite{TitzeHunter07} and \cite{AlzamendiPetersonErathHillmanZanartu21} include reduced-order tissue dynamics, these dynamics are neglected in the present work. Instead, only the corresponding steady-state response is considered. Consequently, the posturing model acts as a static mapping from laryngeal muscle activation levels to the prephonatory configuration parameters $\bar{\varepsilon}$ and $\theta_g$, which are subsequently supplied as inputs to the beam--membrane phonation model (see Fig. \ref{fig:ModelOutline}).

\subsection*{Beam--membrane model}

\begin{figure}[ht!]
    \centering
    \includegraphics[width=0.8\linewidth]{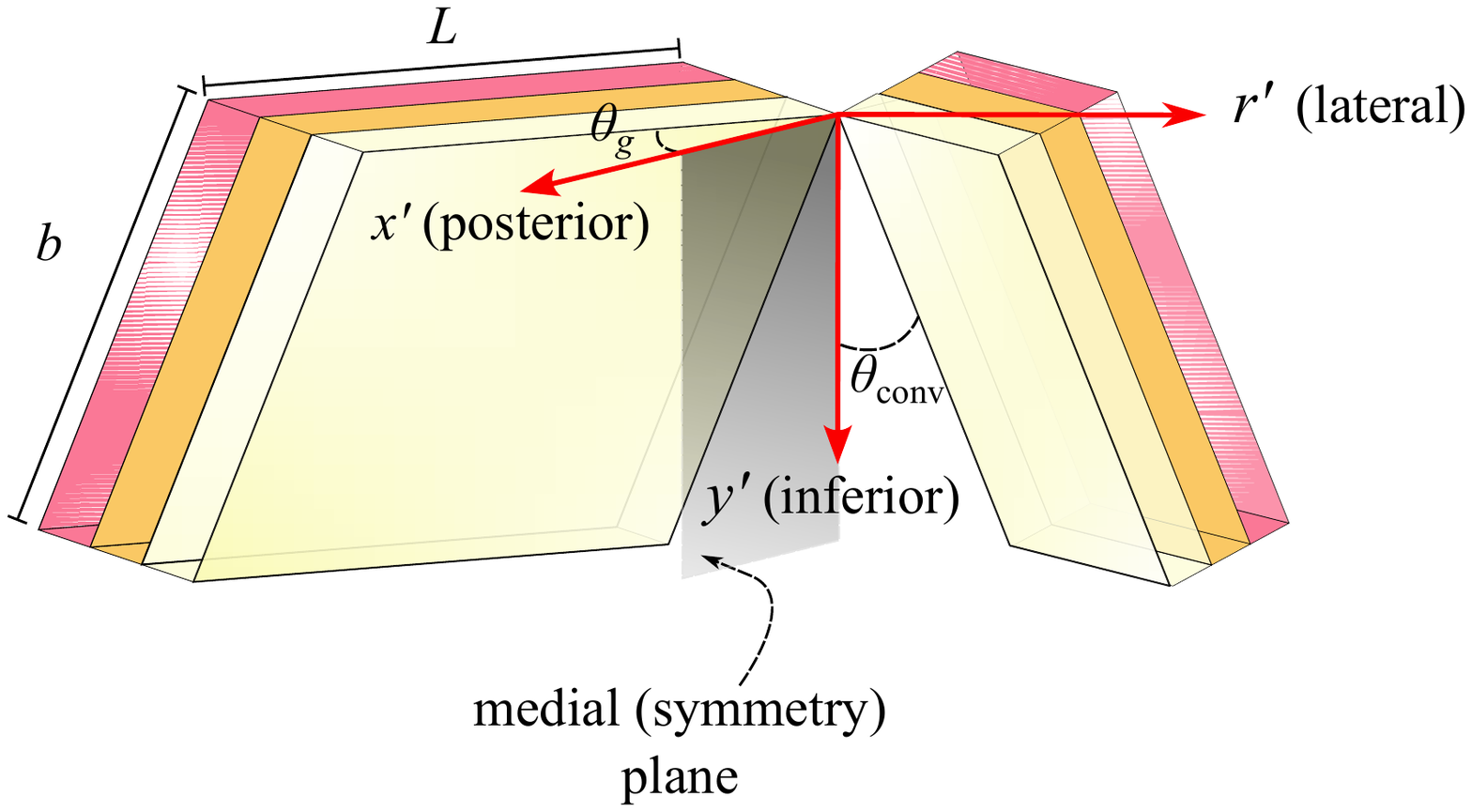}
\caption{
\textbf{Vocal fold configuration and anatomical coordinate system.}
Schematic illustration of the vocal fold geometry and the global coordinate system adopted in the present work. The coordinate axes are aligned with the principal anatomical directions, providing the reference frame used to describe vocal fold kinematics and glottal geometry.
}
    \label{fig:Configuration}
\end{figure}

\begin{figure}[ht!]
    \centering
    \includegraphics[width=0.7\linewidth]{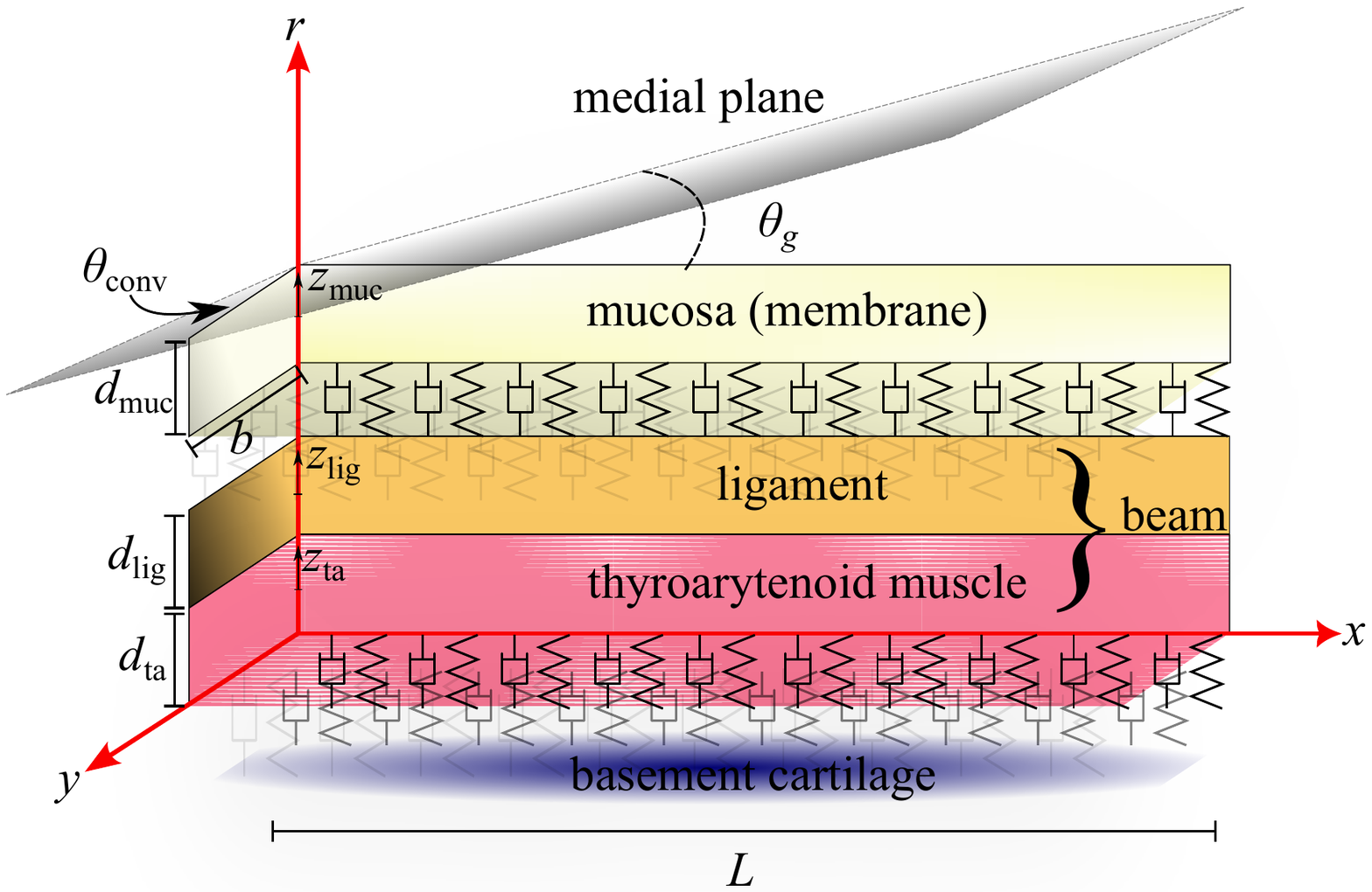}
\caption{
\textbf{Beam--membrane representation of the vocal fold.}
Schematic illustration of the left vocal fold, showing the mucosa, vocal ligament, and thyroarytenoid muscle layers, together with their representation within the proposed beam--membrane framework. The local coordinate system associated with the prestrained configuration is also indicated. The governing equations of the model are derived with respect to this coordinate system.
}
    \label{fig:SchematicMembraneBeam}
\end{figure}

In the proposed modeling framework, each VF is represented as a rectangular prism composed of three anatomical layers, namely the mucosa, the vocal ligament, and the TA muscle, as illustrated in Fig.~\ref{fig:Configuration} and Fig.~\ref{fig:SchematicMembraneBeam}.

The prestrained VF length and inferior--superior width are denoted by $L$ and $b$, respectively, see Fig.~\ref{fig:SchematicMembraneBeam}. In addition to the longitudinal prestrain $\bar{\varepsilon}$ and glottal angle $\theta_g$, the prestrained VF configuration is characterized by a convergence angle $\theta_{\mathrm{conv}}$, defined as the angle between the medial plane and the anterior VF edge, as shown in Fig.~\ref{fig:Configuration} and Fig. ~\ref{fig:SchematicMembraneBeam}. In the present framework, $\theta_{\mathrm{conv}}$ is treated as a user-defined parameter.

For visualization purposes, we introduce a global coordinate system $(x',y',r')$, with its origin located at the intersection of the superior corners of the anterior edges of the vocal folds and its axes aligned with the anatomical directions, as illustrated in Fig.~\ref{fig:Configuration}, wherein
\begin{itemize}
    \item the $x'$-axis points in the posterior direction,
    \item the $y'$-axis points in the inferior direction,
    \item the $r'$-axis points in the lateral direction, with $r'=0$ corresponding to the medial (symmetry) plane.
\end{itemize}

For each VF, we additionally define a local coordinate system $(x,y,r)$ attached to the fold with respect to the \emph{prestrained configuration} as shown in Fig \ref{fig:SchematicMembraneBeam}, wherein
\begin{itemize}
    \item $x \in [0,L]$ denotes the longitudinal coordinate,
    \item $y \in [0,b]$ denotes the inferior--superior coordinate,
    \item $r$ denotes the depth coordinate measured from the base of the TA muscle.
\end{itemize}
The governing equations of the model are formulated with respect to the prestrained configuration.

Each VF is represented using two coupled mechanical components (see Fig.~\ref{fig:SchematicMembraneBeam}):
\begin{itemize}
    \item a one-dimensional  Euler-Bernoulli-type composite beam representing the vocal ligament and TA muscle, with transverse displacement in the $r$-direction denoted by $w_{\mathrm{b}}(x,t)$, and
    \item a two-dimensional membrane representing the mucosal layer, or equivalently the VF medial surface, whose transverse displacement in the $r$-direction is denoted by $w_{\mathrm{m}}(x,y,t)$. Externally, it is loaded by the aerodynamic pressure $p_{\mathrm{aero}}(x,y,t)$, computed using an ideal Bernoulli flow model with viscous corrections \cite{DeckerThomson07}, and by the collision pressure $p_{\mathrm{col}}(x,y,t)$, modeled using a penalty-based contact formulation. Both pressure fields depend implicitly on the instantaneous VF geometry and on the posturing parameters $\theta_g$ and $\theta_{\mathrm{conv}}$.
\end{itemize}

The beam and membrane are mechanically coupled through distributed spring--damper elements that model the viscoelastic interaction between the tissue layers \cite{Zhang16}. In addition, the beam is connected to the basement cartilage through external spring--damper elements that provide elastic support and damping. Herein we assume left-right VF symmetry and thus only the left VF is modeled henceforth. 

In addition to the VF displacement and pressure fields, the dynamic model predicts several quantities of interest, including the minimum glottal area waveform $A_{g,\min}(t)$ and the glottal flow-rate waveform $Q_g(t)$. The radiated acoustic pressure $p_{\mathrm{rad}}(t)$ is estimated from $Q_g(t)$ using the wave-reflection analog (WRA) model illustrated in Fig.~\ref{fig:ModelOutline} and described in Materials and Methods. These outputs are used to compute several acoustic and biomechanical measures, including fundamental frequency $f_{o}$, sound pressure level (SPL), closure quotient (CQ), speed quotient (SQ)\footnote{In the present work, $f_{o}$, CQ, and SQ are estimated from the glottal area waveform $A_{g,\min}(t)$. The closure quotient is defined as the fraction of a phonation cycle during which the glottis remains closed, while the speed quotient is defined as the ratio of the glottal opening time to the glottal closing time.}, and the cycle-averaged maximum collision pressure $p_{\mathrm{col},\max}$.

\section*{Results and discussion}

In this section, we present numerical simulations of the proposed modeling framework to evaluate its predictive capabilities and investigate the influence of glottal geometry on phonatory behavior. The model predictions are compared qualitatively with available computational and experimental observations, and a  parametric study is conducted to examine the role of glottal configuration in shaping VF vibration and collision dynamics.

\subsection*{Simulation setup}
The governing equations of the model (Eqs.~\eqref{eq:MembraneEqnRefined}, \eqref{eq:RotationMembrane}, and \eqref{eq:BeamEqnRefined}), together with their associated boundary conditions, are solved using a finite-difference discretization. One may argue that the spatially discretized form of the proposed framework bears some resemblance to classical multi-mass VF models (see, \textit{e.g.}, \cite{alipour2011mathematical,DejonckereKob09}), though we stress that the present continuum-based model fundamentally differs from lumped-element model frameworks. A principal advantage of the present approach is that its governing parameters are directly related to underlying tissue material properties, while the stiffness and inertial terms are derived primarily from fundamental mechanical principles. Consequently, the proposed framework possesses a stronger physiological foundation and greater physical interpretability than conventional multi-mass models. Furthermore, the proposed framework can transmit bending moments, which is not available in lumped-element models. This capability provides a key biomechanical mechanism through which laryngeal muscle activation can alter VF shape and glottal configuration, thereby enabling the prediction of complex deformation patterns that cannot be captured by the classical multi-mass models. 

In the simulations presented herein, the inferior--superior width $b$ is discretized using $N_y=10$ nodal points (including boundary nodes) and the anterior--posterior length $L$ is discretized using $N_x=20$ nodal points (also including boundary nodes). The supraglottal vocal-tract geometry employed in the acoustic computations corresponds to the vowel $\mathrm{/a/}$ using area functions reported by \cite{Story08}. Temporal discretization uses a time step of $1.7857\times 10^{-6}$ s. This value is selected to ensure compatibility with the WRA acoustic solver, such that pressure waves propagate through each vocal-tract segment in an integer number of time steps. Preliminary spatial and temporal convergence studies were performed to verify that the adopted discretization parameters provide sufficient numerical accuracy.

All simulations are performed over a time interval of $1\,\mathrm{s}$. The reported phonatory measures, including $f_o$, SPL, and glottal flow characteristics, are computed from the final $0.5\,\mathrm{s}$ of each simulation to minimize the influence of transient dynamics. The subglottal pressure is fixed at $P_s=1000\,\mathrm{Pa}$, while the supraglottal pressure is assumed to be zero, such that the transglottal pressure is equal to $P_s$. This transglottal pressure lies within the range reported for male speakers speaking with modal voice \cite[Tables~A2~and~A3]{holmberg1989glottal}. The convergence angle is set to $\theta_{\mathrm{conv}}=0.02\,\mathrm{rad}$, corresponding to a nearly parallel glottal configuration.

To compute static equilibrium configurations, the subglottal pressure is set to zero and the damping coefficients are increased until the transient dynamics decay. The resulting steady-state solutions are then used as the corresponding static VF configurations.

The modeling framework is implemented in Matlab, and the source code is available at \url{https://github.com/UWFluidFlowPhysicsGroup/Membrane-beam-VF-model}. Using the discretization parameters described above, a typical simulation requires less than one minute to complete when executed in MATLAB Online or on a 13th Gen Intel\textsuperscript{\textregistered} Core\texttrademark\ i7-1355U laptop (1.70~GHz). This computational cost is substantially lower than that of high-fidelity phonation models; for example, \cite{xue2012computational} reported that their high-fidelity three-dimensional fluid--structure interaction phonation simulations, performed using a time step of \(3.5\times10^{-6}\,\mathrm{s}\), required approximately \(1200\)  hours on \(128\) processors of a 3.0~GHz parallel computing system to simulate only \(0.126\,\mathrm{s}\) of phonation. This comparison highlights the suitability of the proposed framework for large-scale parametric studies.

\subsection*{Model validation}

\subsubsection*{Static vocal fold configurations}
\begin{figure}[ht!]
    \centering
    \includegraphics[width=0.45\linewidth]{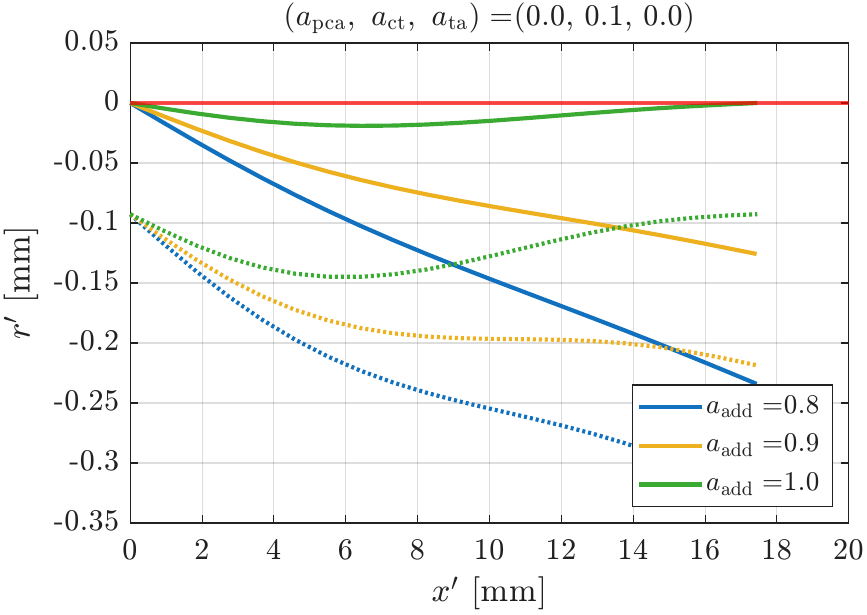}
    \includegraphics[width=0.45\linewidth]{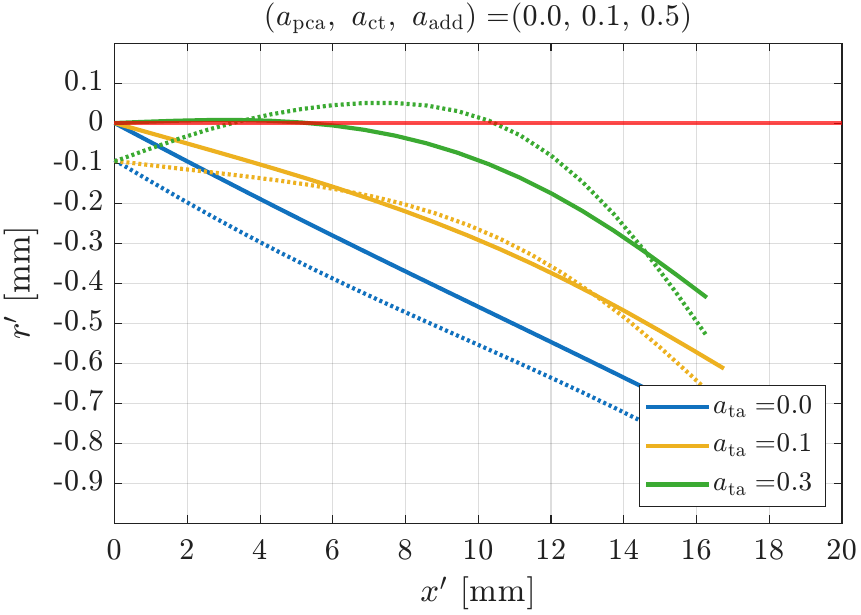}
    \includegraphics[width=0.45\linewidth]{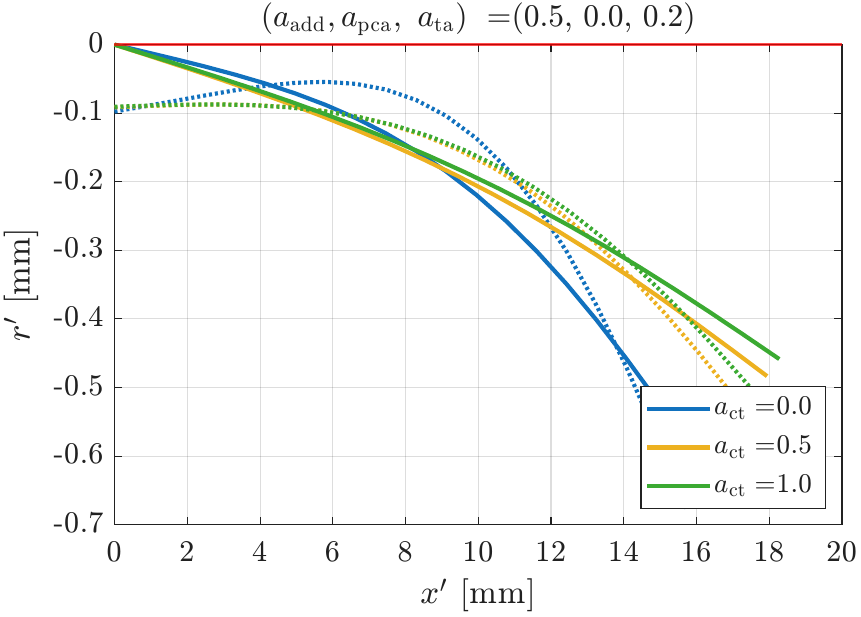}
    \includegraphics[width=0.45\linewidth]{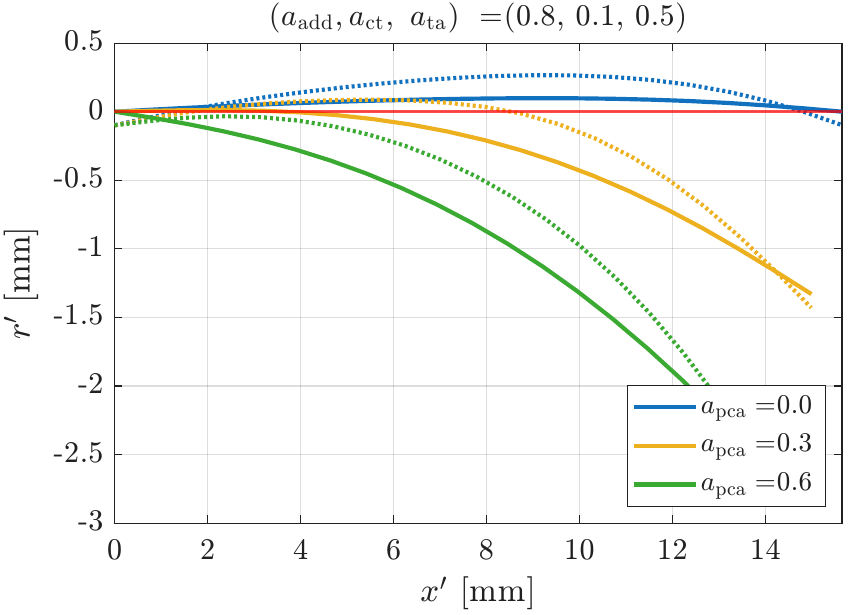}

\caption{
\textbf{Predicted static configurations of the left vocal fold under different muscle activation patterns.}
Predicted static vocal fold shapes corresponding to different intrinsic laryngeal muscle activation levels (solid curves: superior edge; dashed curves: inferior edge). The red line denotes the medial plane. Top-left: effect of the combined adductory muscle activation level, \(a_{\mathrm{add}}=a_{\mathrm{lca}}=a_{\mathrm{ia}}\), with PCA, CT, and TA activation levels held fixed. Top-right: effect of TA activation, \(a_{\mathrm{ta}}\), with adductory, PCA, and CT activation levels held fixed. Bottom-left: effect of CT activation, \(a_{\mathrm{ct}}\), with adductory, PCA, and TA activation levels held fixed. Bottom-right: effect of PCA activation, \(a_{\mathrm{pca}}\), with adductory, CT, and TA activation levels held fixed.
}
    \label{fig:StaticShapes}
\end{figure}

We first demonstrate that the proposed framework is capable of reproducing physiologically relevant static VF configurations that are qualitatively consistent with clinical observations and previous high-fidelity computational studies. Figure~\ref{fig:StaticShapes} illustrates the predicted static VF shapes for several representative combinations of intrinsic laryngeal muscle activations.

The top-left panel illustrates the influence of the primary adductory muscles, represented by the combined activation level $a_{\mathrm{add}}=a_{\mathrm{lca}}=a_{\mathrm{ia}}$, with zero TA activation and a fixed CT activation level of $0.1$. As expected, increasing adductory muscle activation progressively reduces the posterior glottal gap and eventually promotes posterior glottal closure. Because TA activation remains low, the resulting glottal conformations display anterior glottal openings, a clinically observed pattern associated with incomplete VF approximation in the anterior portion of the glottis \cite{ChoiYeBerkeKreiman93,ChhetriNeubauer15}.

The top-right panel illustrates the influence of TA activation while maintaining the CT activation level at $0.1$ and the combined adductory muscle activation at $a_{\mathrm{add}}=0.5$. Increasing $a_{\mathrm{ta}}$ produces progressive medial bulging of the VF body due to bending moments arising from the nonuniform stress distribution across the different tissue layers of the folds \cite{serry2023euler}. This medial bulging promotes anterior and mid-membranous glottal closure. Because the adductory activation level is relatively modest, the increased medial curvature induced by the TA muscle gives rise to a concave glottal configuration accompanied by a persistent posterior glottal opening, consistent with experimental observations \cite{ChoiYeBerkeKreiman93}.

Comparing the superior and inferior VF edges in the top-right panel, we see that near the mid-membranous region, increasing TA activation causes the inferior edge to move medially ahead of the superior edge, resulting in a divergent glottal configuration. This effect is largely absent near the anterior and posterior ends, where the VF motion is constrained by the boundary conditions. Within the proposed beam--membrane framework this behavior arises naturally from the coupling between axial loading, bending, and the layered tissue structure. Specifically, TA activation induces bending that drives the medial surface outward. Because the inferior portion of the fold is more compliant than the superior portion, it undergoes greater medial displacement, thereby producing an inferior-edge lead. Interestingly, muscle-induced inferior-edge lead has previously been incorporated into lumped-mass phonation models through empirical convergence-angle rules; whereas lumped-element models must impose this in an \textit{ad hoc} manner \cite{TitzeStory02}, this outcome arises naturally in the present model.  

The bottom-left panel illustrates the effect of increasing CT activation for fixed values of $a_{\mathrm{add}}=0.5$ and $a_{\mathrm{ta}}=0.2$. Increasing $a_{\mathrm{ct}}$ elongates and stiffens the VFs, thereby counteracting the medial bulging induced by TA activation. As a result, the folds become progressively straighter, exhibiting reduced medial curvature and a more elongated configuration. This behavior is consistent with biomechanical and experimental observations demonstrating that CT activation increases longitudinal tension while reducing VF convexity \cite{ChhetriNeubauerBerry12}.

The bottom-right panel illustrates the influence of PCA activation for fixed values of $a_{\mathrm{add}}=0.8$ and $a_{\mathrm{ta}}=0.5$. Starting from a fully closed glottal configuration produced by the combined action of the adductory and TA muscles, increasing PCA activation progressively abducts the posterior portion of the VFs, leading to the formation of a posterior glottal opening. This behavior is consistent with the well-established role of the PCA muscle as the primary VF abductor and agrees with clinical observations \cite{MorrisonRammage93}. We note that the slight overlap of the medial surfaces visible in some predicted configurations is a consequence of the penalty-based contact formulation employed in the model, in which interpenetration is permitted but penalized through the contact force law.

Overall, the model naturally reproduces a broad range of clinically observed glottal configurations, including anterior glottal opening, posterior glottal opening, and near-complete closure. These predictions are in qualitative agreement with the experimental, clinical, and computational studies discussed in the Introduction. The ability of the proposed framework to reproduce these characteristic static configurations provides confidence that the underlying posturing mechanics capture the principal biomechanical mechanisms governing VF geometry and motivates the subsequent investigation of the corresponding phonatory dynamics.

\subsubsection*{Exemplar case of sustained modal phonation}
To illustrate the dynamic behavior of the proposed model during sustained modal phonation, we consider a representative simulation with
\[
(a_{\mathrm{lca}},
a_{\mathrm{ia}},
a_{\mathrm{pca}}, a_{\mathrm{ct}},
a_{\mathrm{ta}}
)
=
(0.8,0.8,0,0.3,0.4).
\]
This activation pattern corresponds to strong adductory muscle activation, no PCA activation, and moderate CT and TA activation. Such a configuration promotes complete glottal closure while maintaining sufficient longitudinal tension and tissue stiffness to support self-sustained oscillation. 

\begin{figure}[ht!]
    \centering
    $$
    (a_{\mathrm{lca}},
    a_{\mathrm{ia}},
    a_{\mathrm{pca}}, a_{\mathrm{ct}},
    a_{\mathrm{ta}},)=
    (0.8,0.8,0,0.3,0.4)$$

    \includegraphics[width=0.425\linewidth]{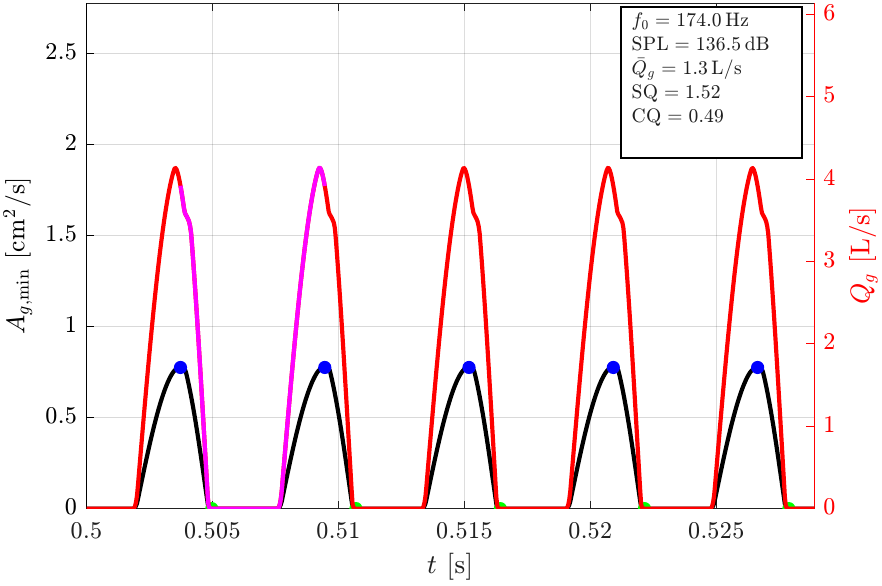}~
    \includegraphics[width=0.4\linewidth]{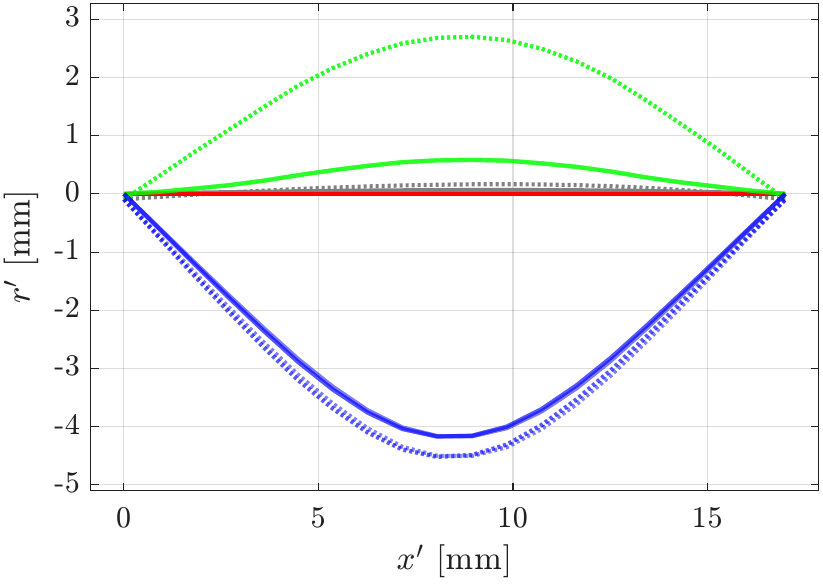}

    \includegraphics[width=0.8\linewidth]{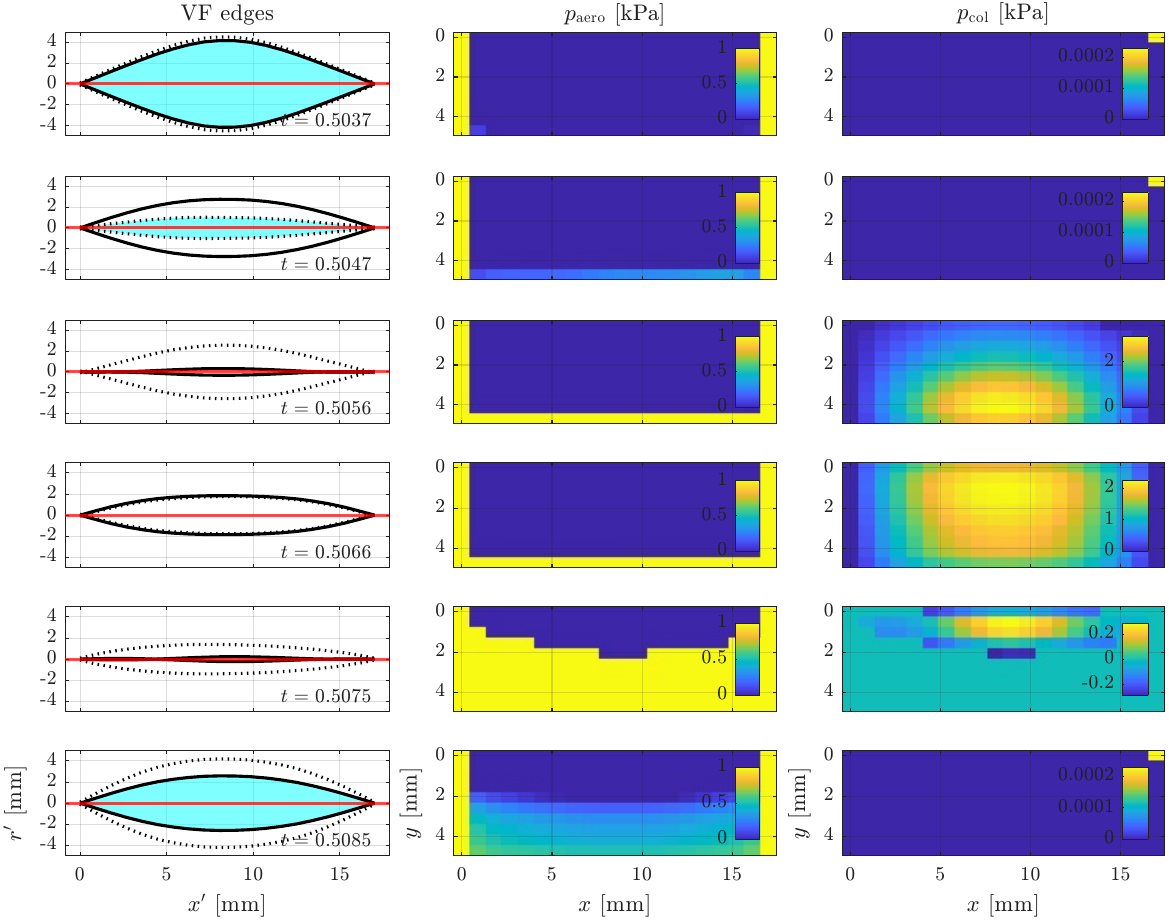}

    \caption{
    \textbf{Representative sustained modal phonation simulation.}
Top-left: glottal area and glottal flow-rate waveforms. The magenta-highlighted segment identifies the oscillation cycle illustrated in the lower panels. Top-right: left vocal fold medial-surface profiles corresponding to maximum opening (blue), maximum closure (green), and the static equilibrium configuration (gray). The maximum-opening and maximum-closure profiles correspond to the highlighted peak and nadir of the glottal area waveform, respectively. Solid and dashed curves denote the superior and inferior vocal fold edges, respectively, while the red line indicates the medial plane. Bottom: evolution of the vocal fold and glottal configuration, aerodynamic pressure distribution, and contact-pressure distribution over one representative oscillation cycle. The first column of panels shows the medial-surface shapes of both vocal folds, with the glottal region shaded in cyan. The disappearance of the cyan region indicates complete glottal closure. The second and third columns show the corresponding aerodynamic and contact-pressure distributions, respectively.
   }
   \label{fig:Example}
\end{figure}

Figure~\ref{fig:Example} presents the resulting phonatory dynamics. The top-left panel shows the glottal area and glottal flow-rate waveforms, the top-right panel depicts the static  configuration of the left VF together with the shapes corresponding to maximum and minimum value of the glottal area waveform, and the bottom panels illustrate the evolution of the VF medial surface and glottal shape and the associated aerodynamic and contact pressure distributions over one oscillation cycle. The highlighted segment of the waveform identifies the cycle displayed in the lower panels.

The glottal area waveform exhibits the periodic open--closed behavior characteristic of sustained phonation. The opening phase occupies a larger fraction of the cycle than the closing phase, yielding a speed quotient of $\mathrm{SQ}=1.52$. This indicates that glottal closure occurs more rapidly than glottal opening, a characteristic commonly associated with healthy modal phonation \cite{chen2002electroglottographic}. Reported speed quotient values for normal modal phonation typically range from approximately \(1.3\) to \(2.8\), depending on factors such as sex, vowel production, and the measurement methodology employed \cite{chen2002electroglottographic}. The closure quotient is $\mathrm{CQ}=0.49$, indicating that the glottis remains closed for approximately half of the oscillation period. Such a value is consistent with modal phonation, in which the open and closed phases occupy comparable portions of the cycle (typical mean value of CQ  in healthy adults is approximately 0.50 \cite{mehta2010voice}).

The glottal flow-rate waveform follows the same general periodic pattern as the glottal area waveform, increasing during glottal opening and decreasing during closure. However, the flow waveform is not simply a scaled version of the glottal area waveform. In particular, the flow rate exhibits a slight decrease near the end of the opening phase and a nonuniform decline during closure, indicating that the flow dynamics are not simply proportional to the glottal area. Similar waveform characteristics have been observed in both high-fidelity computational simulations \cite[Fig.~6(b)]{JiangGengZhengXue24} and reduced-order phonation models \cite[Fig.~6(b) and Fig.~8(b)]{StoryTitze95}.

The top-right panel further illustrates the influence of dynamic effects on VF configuration. The static equilibrium shape corresponds to a nearly flat configuration with complete glottal closure, whereas the shapes attained during oscillation exhibit substantially greater convexity/concavity. In particular, the glottis assumes a convergent profile near maximum opening and a divergent profile near maximum closure, highlighting the dominant role of fluid--structure interaction in shaping the glottal geometry during phonation. 

The lower panels of Fig.~\ref{fig:Example} demonstrate the spatiotemporal evolution of the medial-surface deformation and pressure fields during the selected cycle. During the opening phase, the glottis is predominantly convergent, whereas during the closing phase it becomes predominantly divergent. This alternating convergent--divergent pattern is consistent with clinically observed mucosal-wave-like motion and with established descriptions of phonatory vibration (see, \textit{e.g.}, \cite{JiangLinHanson00}). The phase difference between the inferior and superior portions of the VF surface contributes to the asymmetric pressure distribution over the cycle and is essential for aerodynamic energy transfer to the tissue \cite{Titze88,ThomsonMongeauFrankel05}.

The aerodynamic pressure distribution further illustrates the mechanism sustaining self-oscillation. During the opening phase, the pressure acting on the medial surface is predominantly positive, thereby assisting the outward motion of the VFs. During the closing phase, the aerodynamic pressure magnitude is substantially reduced over much of the medial surface, except near regions of local flow obstruction, particularly at the anterior and posterior boundaries. This cycle-dependent pressure asymmetry produces a net positive transfer of energy from the airflow to the VF tissue and is consistent with the classical myoelastic theory of self-sustained phonation \cite{Titze88, JiangLinHanson00}.

Heightened contact pressure has been implicated as a factor precipitating phonatory vocal hyperfunction \cite{HillmanSteppVanStanZanartuMehta20} and is thus a critical monitoring parameter for vocal health. Collision pressure, both magnitude and location, are natural outcomes of the present model. During the closing phase, contact is initiated near the inferior edge and subsequently propagates toward the superior edge, reflecting the vertical phase difference in VF motion. This progression is clearly visible in the pressure fields at the selected time instants during closure. As the VFs begin to separate, the contact model predicts negative contact-pressure values arising from the damping component of the contact law. Although not intended to represent adhesion explicitly, this behavior qualitatively resembles the adhesive forces incorporated in high-fidelity phonation simulations to model the resistance of vocal fold tissues to separation following contact \cite{BhattacharyaSiegmund15}.

Broadly, these results highlight the ability of the proposed framework to resolve physiologically relevant contact patterns during phonation. While such patterns have previously been investigated using simplified theoretical analyses based on the eigenfunctions of two-dimensional wave equations \cite{SmithTitze18}, the present framework allows them to emerge naturally from the coupled structural, aerodynamic, and contact dynamics. This capability facilitates the investigation of both the spatial and temporal characteristics of vocal fold collision under physiologically realistic conditions.

Overall, the representative simulation demonstrates that the proposed framework can reproduce several qualitative features of sustained modal phonation, including periodic glottal opening and closure, asymmetric glottal area and flow waveforms, alternating convergent and divergent glottal shapes from mucosal wave propagation, phase-dependent aerodynamic pressure asymmetry, and propagating contact along the inferior--superior direction. These features support the suitability of the model for investigating how static glottal configuration and muscle activation influence phonatory dynamics.

\subsection*{Phonatory measures as functions of CT and TA activation}
Movahhedi \textit{et al.} \cite{MovahhediGengXueZheng21} and Jiang \textit{et al.} \cite{JiangGengZhengXue24} presented high-fidelity finite-element simulation results of the influence of intrinsic muscle activation on VF posturing and phonation measures, which serves herein as a comparator for the qualitative validation of the present framework. Akin to their studies, we consider activation combinations satisfying $a_{\mathrm{ta}} \in [0,1]$ and $a_{\mathrm{ct}} \in [0,1]$, and generate activation maps for several phonatory measures. The biomechanical effects of these muscles have also been investigated empirically \cite{ChhetriNeubauerSoferBerry14}.

\begin{figure}[ht!]
    \centering
    \includegraphics[width=0.49\linewidth]{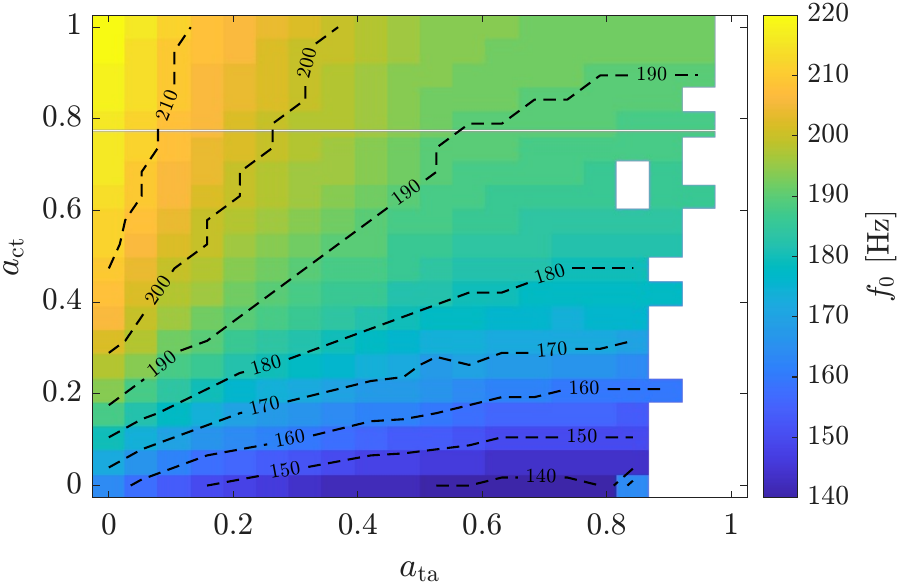}
    \includegraphics[width=0.49\linewidth]{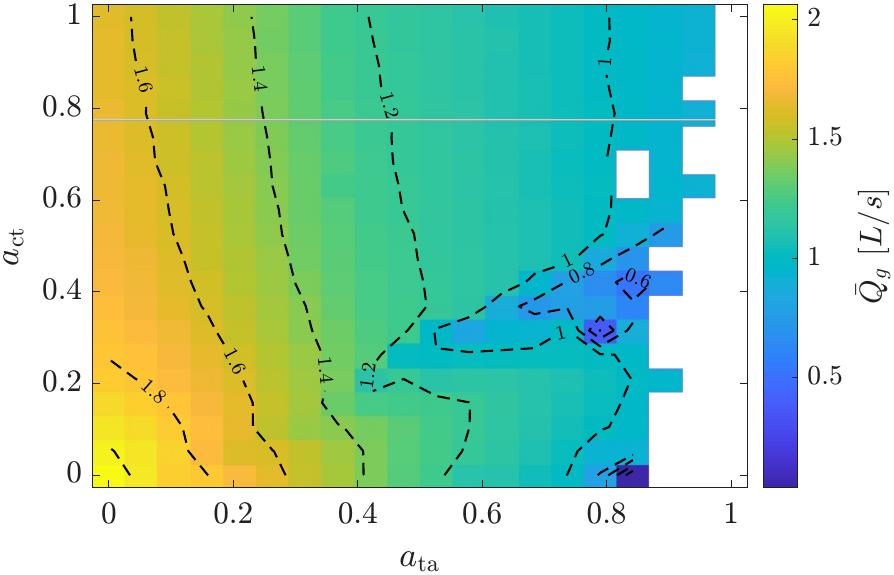}
\includegraphics[width=0.49\linewidth]{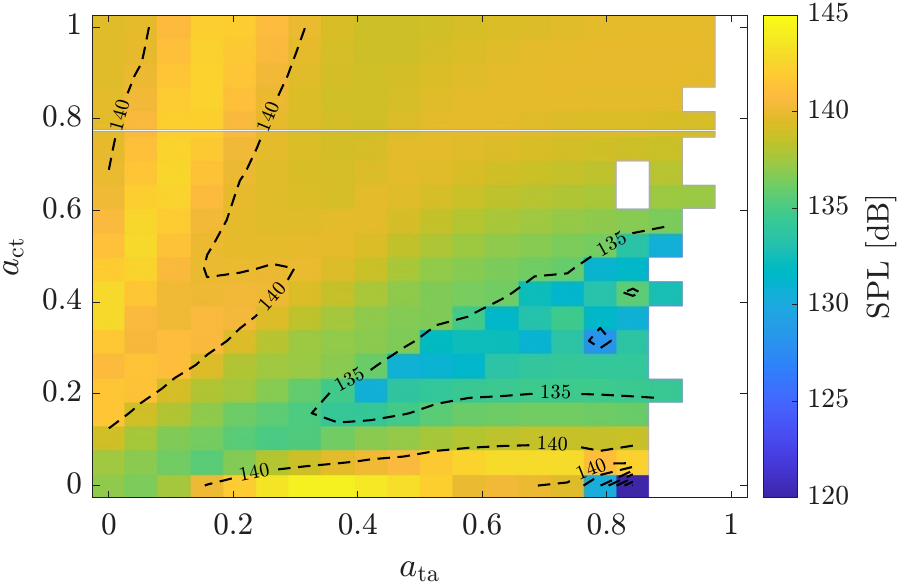}
    \includegraphics[width=0.49\linewidth]{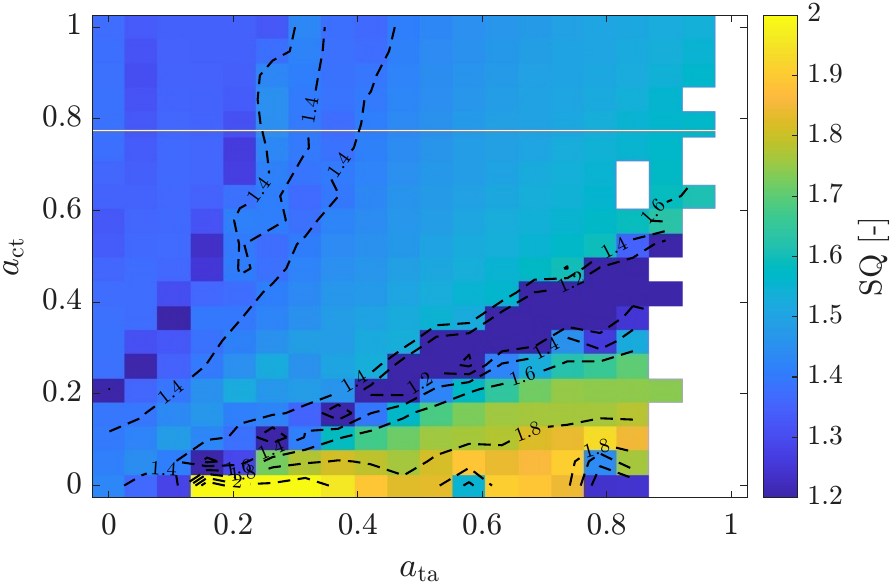}
\includegraphics[width=0.49\linewidth]{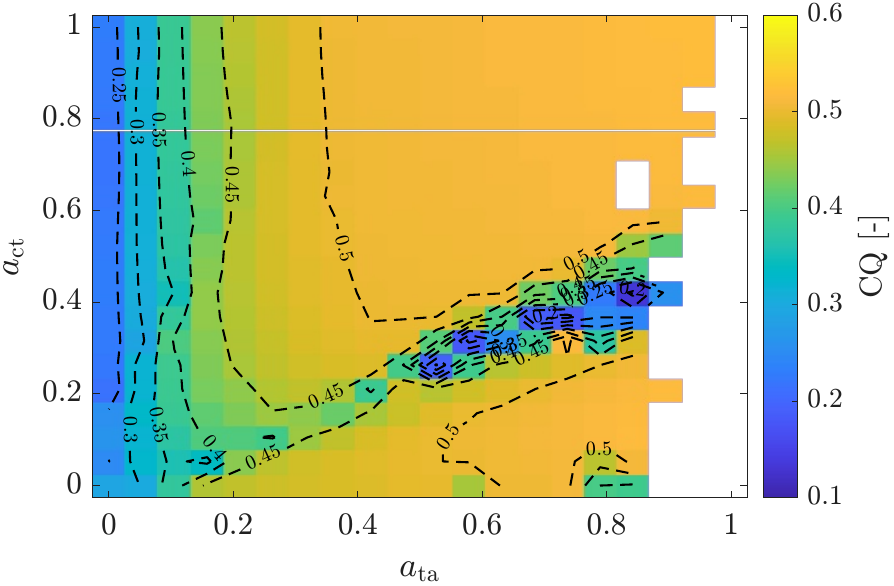}
\caption{ \textbf{Influence of CT and TA activation on phonatory measures.} Activation maps of selected phonatory measures as functions of CT activation $a_{\mathrm{ct}}$ and TA activation $a_{\mathrm{ta}}$ for fixed adductory activation $a_{\mathrm{lca}}=a_{\mathrm{ia}}=0.6$ and no PCA activation, $a_{\mathrm{pca}}=0$. The plotted quantities include fundamental frequency $f_{o}$, average glottal flow rate $\bar{Q}_{g}$, sound pressure level SPL, speed quotient SQ, and closure quotient CQ. Blank regions indicate activation combinations for which physiologically relevant sustained oscillations were not obtained.}
    \label{fig:CTvsTA}
\end{figure}

Figure~\ref{fig:CTvsTA} presents the resulting activation maps for fixed, moderate adductory muscle activation levels, $a_{\mathrm{lca}}=a_{\mathrm{ia}}=0.6$, and zero PCA activation ($a_{\mathrm{pca}}=0$). The quantities shown are analogous to those reported in Fig.~8 of \cite{MovahhediGengXueZheng21} and Fig.~5 of \cite{JiangGengZhengXue24}. It should be noted, however, that the models employed in \cite{MovahhediGengXueZheng21,JiangGengZhengXue24} do not explicitly account for the effects of the adductory (LCA and IA) and abductory (PCA) muscles on VF posture and phonation. Instead, vocal fold adduction is prescribed directly through the model configuration. In contrast, the present framework incorporates these muscles explicitly, allowing the glottal posture to emerge naturally from the underlying muscle activation pattern.

The blank regions in Fig.~\ref{fig:CTvsTA} correspond to muscle activation combinations for which sustained oscillations were not observed. These regions occur primarily at high levels of TA activation wherein the shortening action of the TA muscle results in the longitudinal membrane tension becoming insufficient to sustain self-oscillation, leading to the loss of stable phonatory behavior. Similar regions of phonation cessation have been reported in both high-fidelity and lumped-mass phonation models for certain muscle activation combinations, including cases with elevated TA activation levels (see, e.g., \cite[Fig.~7]{AlzamendiPetersonErathHillmanZanartu21}, \cite[Fig.~8]{MovahhediGengXueZheng21}, and \cite[Fig.~5]{JiangGengZhengXue24}).

The fundamental frequency map demonstrates that $f_o$ is governed predominantly by CT activation. Increasing $a_{\mathrm{ct}}$ produces a substantial increase in $f_o$, whereas increasing TA activation generally decreases $f_o$. This trend is consistent with the established physiological role of the CT muscle, which elongates the VFs and increases their longitudinal tension. In contrast, under the present activation conditions, TA activation tends to shorten and thicken the VFs, thereby reducing the effective longitudinal tension. The competing actions of the CT and TA muscles therefore give rise to the observed variation in fundamental frequency.

The contour lines provide additional insight into the sensitivity of $f_o$ to muscle activation. In particular, the contour slopes increase with increasing CT activation, indicating that progressively larger changes in $a_{\mathrm{ct}}$ are required to achieve the same increase in $f_o$. This suggests that the sensitivity of $f_o$ to CT activation decreases at higher activation levels. The predicted increase in $f_o$ with CT activation, together with the reduction in sensitivity at high CT activation levels, is consistent with trends reported by both lumped-mass models \cite[Fig.~8]{AlzamendiPetersonErathHillmanZanartu21} and high-fidelity continuum models \cite[Fig.~8(a)]{MovahhediGengXueZheng21}.

A subtle difference between the present predictions and those reported in some previous studies is that TA activation decreases $f_o$ monotonically throughout the phonatory regime. In contrast, several existing models in addition to experimental studies predict an initial increase in $f_o$ with TA activation at low TA activation levels, followed by a decrease as TA activation is increased further \cite{MovahhediGengXueZheng21,AlzamendiPetersonErathHillmanZanartu21,ChhetriNeubauerSoferBerry14}. Nevertheless, the behavior at moderate and high TA activation levels is qualitatively similar to that predicted here. The trends reported by \cite[Fig.~5(a)]{JiangGengZhengXue24} exhibit some similarities to the present results but also display additional nonlinear features, making monotonic relationships less apparent across portions of the activation map.

The proposed model predicts fundamental frequencies in the approximate range of $140$--$220\,\mathrm{Hz}$, which is comparable to the ranges reported by \cite[Fig.~8(a)]{MovahhediGengXueZheng21} ($150$--$190\,\mathrm{Hz}$) and \cite[Fig.~5(a)]{JiangGengZhengXue24} ($100$--$220\,\mathrm{Hz}$). It is worth noting that neither the present framework nor these high-fidelity models incorporate two-way acoustic coupling with the vocal tract. In contrast, lumped-mass models that include two-way source--tract interaction predict substantially broader frequency ranges \cite[Fig.~8]{AlzamendiPetersonErathHillmanZanartu21}, extending from approximately $100$ to $300\,\mathrm{Hz}$. Clinical studies have reported that the fundamental frequency $f_{o}$ can vary over a wide range, approximately $100$--$400\,\mathrm{Hz}$, depending on the levels of CT and TA muscle activation \cite{martinez2025toward}.

The mean glottal flow rate $\bar{Q}_{g}$ generally decreases with increasing TA activation. This behavior is associated with the increased medial bulging produced by the TA muscle, which promotes glottal closure and reduces the effective flow area available during phonation. By comparison, variations in CT activation have a relatively modest influence on $\bar{Q}_{g}$ in general. Overall, these trends are in qualitative agreement with the predictions of the high-fidelity computational model presented in \cite[Fig.~8(d)]{MovahhediGengXueZheng21}.

The SPL map exhibits a strongly nonlinear dependence on CT and TA activation, characterized by a cone-shaped region of low SPL that separates two regions of elevated acoustic output. Nevertheless, larger SPL values tend to occur at higher CT activation levels, particularly when TA activation is relatively small. These trends are broadly consistent with the predictions of high-fidelity computational models \cite[Fig.~8(e)]{MovahhediGengXueZheng21} and \cite[Fig.~5(c)]{JiangGengZhengXue24}. In particular, the overall organization of the contour lines predicted by the present model is qualitatively similar to that reported in \cite[Fig.~5(c)]{JiangGengZhengXue24}, suggesting that the proposed framework captures the primary mechanisms governing SPL variation across the muscle-activation space.

The speed quotient map also exhibits a highly nonlinear dependence on CT and TA activation, with contour patterns that bear some resemblance to those observed in the $f_o$ map. The largest values of $SQ$ generally occur at low CT activation levels, indicating phonatory cycles characterized by relatively slow opening phases and more rapid closure phases. As CT activation increases, the speed quotient tends to decrease, reflecting a more symmetric temporal distribution between the opening and closing phases of the oscillation cycle. The overall contour structure, as well as the concentration of the highest $SQ$ values at low CT activation levels, is consistent with the predictions of high-fidelity simulations reported in \cite[Fig.~8(b)]{MovahhediGengXueZheng21}.

The closure-quotient map exhibits a highly nonlinear dependence on CT and TA activation. Two prominent regions of low CQ values are observed: one extending radially along the $a_{\mathrm{ta}}=0$ axis and a second extending diagonally from the origin toward approximately $(a_{\mathrm{ta}},a_{\mathrm{ct}})=(0.8,0.4)$. Between these regions lie two zones of elevated CQ values. Overall, CQ increases with increasing TA activation, reflecting the enhanced medial bulging and glottal closure produced by the TA muscle. As the folds remain in contact for a larger fraction of each oscillation cycle, the duration of the closed phase increases, leading to higher closure quotients. The low-CQ region at zero TA activation and the elevated CQ values at low CT activation levels are in qualitative agreement with the high-fidelity predictions reported in \cite[Fig.~8(c)]{MovahhediGengXueZheng21}.

Overall, the activation maps demonstrate that the proposed framework reproduces several physiologically meaningful trends observed in high-fidelity computational models. In particular, the model captures the dominant influence of CT activation on fundamental frequency, the reduction in average flow rate associated with increasing TA activation, the nonlinear dependence of SPL on muscle activation, and the characteristic variations of speed and closure quotients across the activation space. These results provide additional evidence that the proposed framework captures the principal biomechanical mechanisms governing phonation while offering substantially lower computational cost than high-fidelity fluid--structure interaction models.

\subsection*{Abductory effects of PCA activation on phonation}
In this section, we present a brief parametric study illustrating the capability of the proposed framework to provide biomechanical insights into voice production. As demonstrated in the previous section, activation of the PCA muscle induces posterior glottal opening by abducting the posterior portion of the VFs. Here, we investigate how increasing PCA activation influences phonatory behavior, with particular emphasis on its effects on glottal configuration and VF collision dynamics.

\subsubsection*{Exemplar case with posterior glottal opening}
We consider here an exemplar case obtained by increasing PCA activation relative to the modal-phonation case discussed previously:
\[
(a_{\mathrm{lca}},
a_{\mathrm{ia}},
a_{\mathrm{pca}},
a_{\mathrm{ct}},
a_{\mathrm{ta}})
=
(0.8,0.8,0.5,0.3,0.4).
\]
This activation pattern differs from the modal case only through the increased PCA activation, which induces a persistent posterior glottal opening while maintaining substantial adductory muscle activation.

\begin{figure}[ht!]
 \centering
$$
(a_{\mathrm{lca}},
a_{\mathrm{ia}},
a_{\mathrm{pca}}, a_{\mathrm{ct}},
a_{\mathrm{ta}},
)
=
(0.8,0.8,0.5,0.3,0.4)
$$
\includegraphics[width=0.425\linewidth]{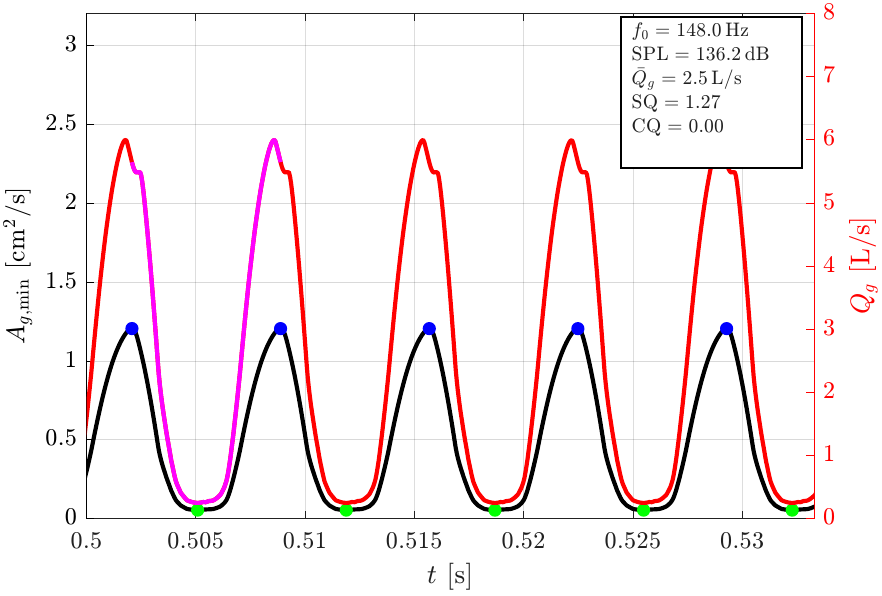}~\includegraphics[width=0.4\linewidth]{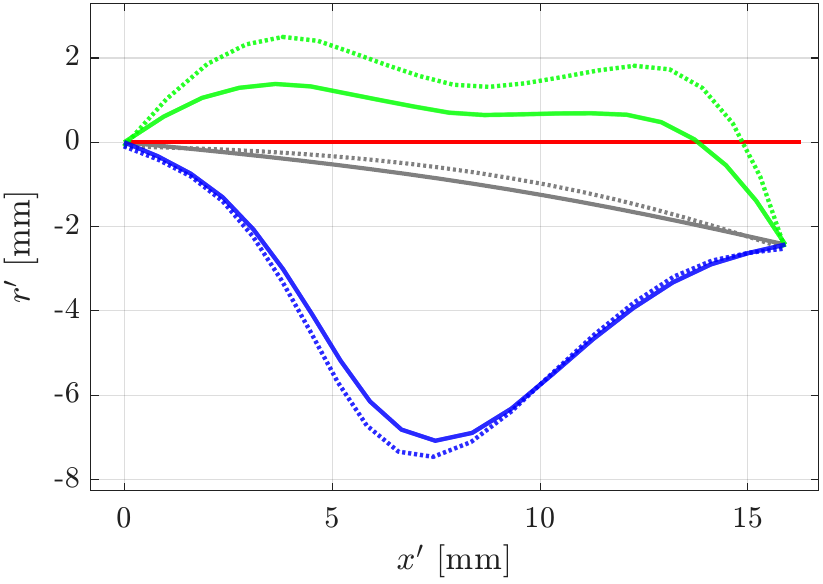}
\includegraphics[width=0.8\linewidth]{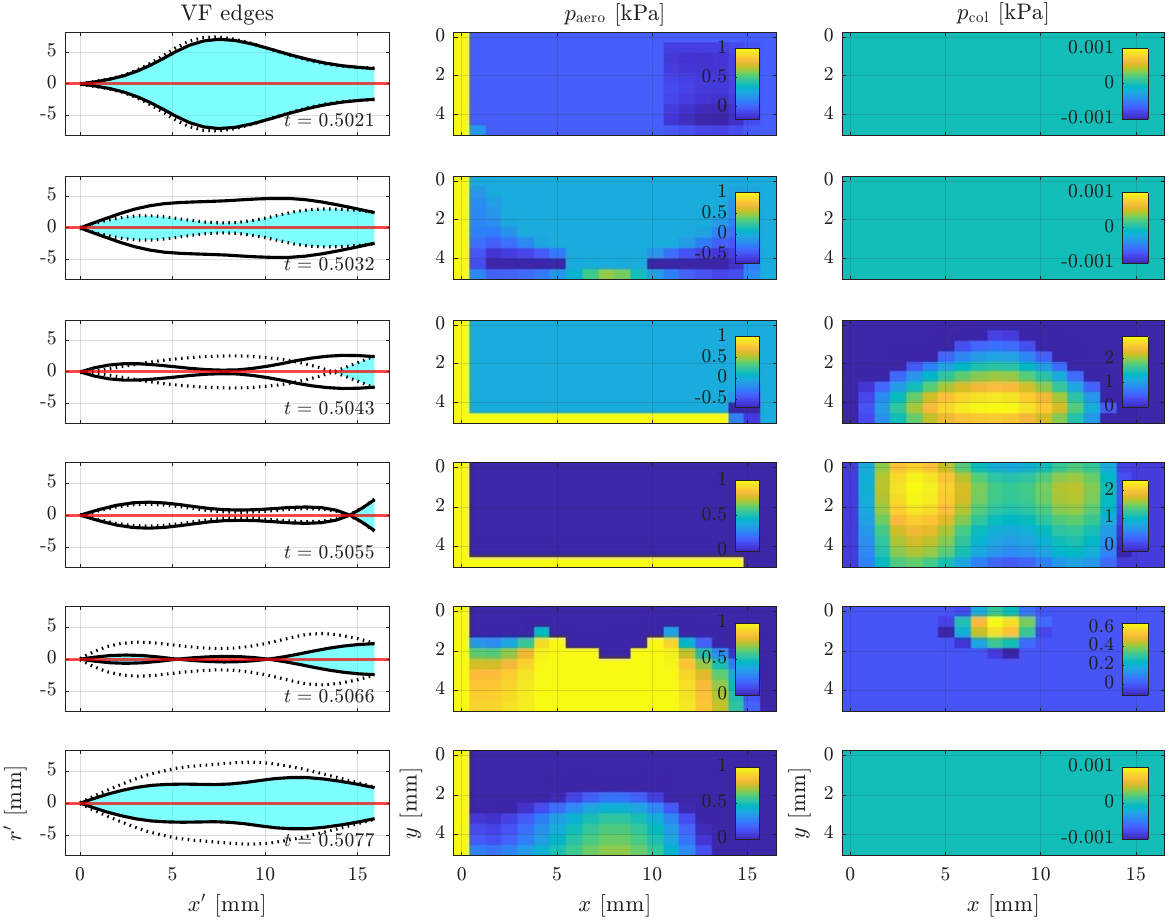}

\caption{
\textbf{Sustained phonation simulation with posterior glottal opening.}
Top-left: glottal area and glottal flow-rate waveforms. The magenta-highlighted segment identifies the oscillation cycle illustrated in the lower panels. Top-right: left vocal fold medial-surface profiles corresponding to maximum opening (blue), maximum closure (green), and the static equilibrium configuration (gray). The maximum-opening and maximum-closure profiles correspond to the highlighted peak and nadir of the glottal area waveform, respectively. Solid and dashed curves denote the superior and inferior vocal fold edges, respectively, while the red line indicates the medial plane. Bottom: evolution of the vocal fold and glottal configuration, aerodynamic pressure distribution, and contact-pressure distribution over one representative oscillation cycle. The first column of panels shows the medial-surface shapes of both vocal folds, with the glottal region shaded in cyan. The disappearance of the cyan region indicates complete glottal closure. The second and third columns show the corresponding aerodynamic and contact-pressure distributions, respectively.
}
   \label{fig:Example2}
\end{figure}

As in Fig.~\ref{fig:Example}, Fig.~\ref{fig:Example2} presents the glottal area and glottal flow-rate waveforms, representative VF configurations at maximum opening and maximum closure, and the evolution of the VF medial surface together with the corresponding aerodynamic and contact pressure distributions over one oscillation cycle. Similar to the modal case, the glottal area and flow-rate waveforms remain periodic, and the glottis exhibits a predominantly convergent configuration during opening and a predominantly divergent configuration during closure. These observations indicate that the fundamental fluid--structure interaction mechanism responsible for self-sustained oscillation is preserved despite the presence of a posterior glottal gap.

The VF configurations attained during phonation differ substantially from the corresponding static equilibrium configuration obtained in the absence of subglottal pressure. In particular, the dynamic configurations exhibit pronounced nonlinear curvature and significantly larger deformation amplitudes, highlighting the important role of aerodynamic loading and structural dynamics in shaping the glottal geometry during phonation.

The figure further illustrates the spatiotemporal evolution of the aerodynamic and contact pressure fields. Comparison with the modal phonation case shown in Fig.~\ref{fig:Example} reveals that the posterior glottal opening substantially alters the spatial distribution of both aerodynamic and contact pressures throughout the oscillation cycle. In particular, the pressure fields become markedly asymmetric in the anterior--posterior direction, reflecting the nonuniform glottal geometry introduced by PCA activation. Collision pressure exhibits a peak at the anterior-posterior midpoint that then moves anteriorly later in the cycle. These results demonstrate that posterior glottal opening influences not only the mean glottal configuration but also the spatial characteristics of the fluid--structure interaction and contact mechanics governing phonation.

\subsubsection*{Contact forces and posterior glottal opening}
The exemplar case presented in the previous section illustrates how posterior glottal opening induced by PCA activation alters phonatory behavior, particularly through the introduction of pronounced anterior--posterior asymmetries in the aerodynamic and contact pressure distributions. Since elevated collision forces have been associated with VF trauma and phonotraumatic disorders \cite{HillmanSteppVanStanZanartuMehta20}, we exercise the model to investigate the role of posterior glottal opening on VF collision dynamics.

To this end, we examine the effect of PCA activation on the cycle-averaged maximum contact pressure  $p_{\mathrm{col,max}}$, together with the corresponding anterior--posterior location at which it occurs. Specifically, we consider the reference case
\[
(a_{\mathrm{lca}},a_{\mathrm{ia}},a_{\mathrm{pca}},a_{\mathrm{ct}},a_{\mathrm{ta}})
=
(0.8,0.8,0,0.3,0.4),
\]
which corresponds to near-complete glottal closure, and progressively increase the PCA activation level from $a_{\mathrm{pca}}=0$ to $a_{\mathrm{pca}}=1$ while holding all other muscle activations fixed.

\begin{figure}[ht!] 
    \centering $$ (a_{\mathrm{lca}},a_{\mathrm{ia}},a_\mathrm{ct},a_\mathrm{ta})=(0.8,0.8,0.3,0.4) $$ 
    \includegraphics[width=0.5\linewidth]{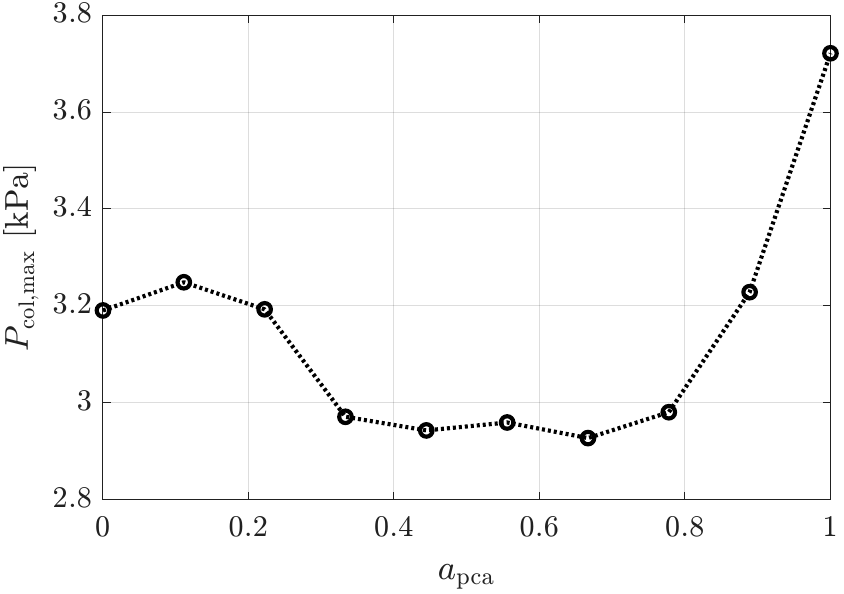}    
    \includegraphics[width=0.49\linewidth]{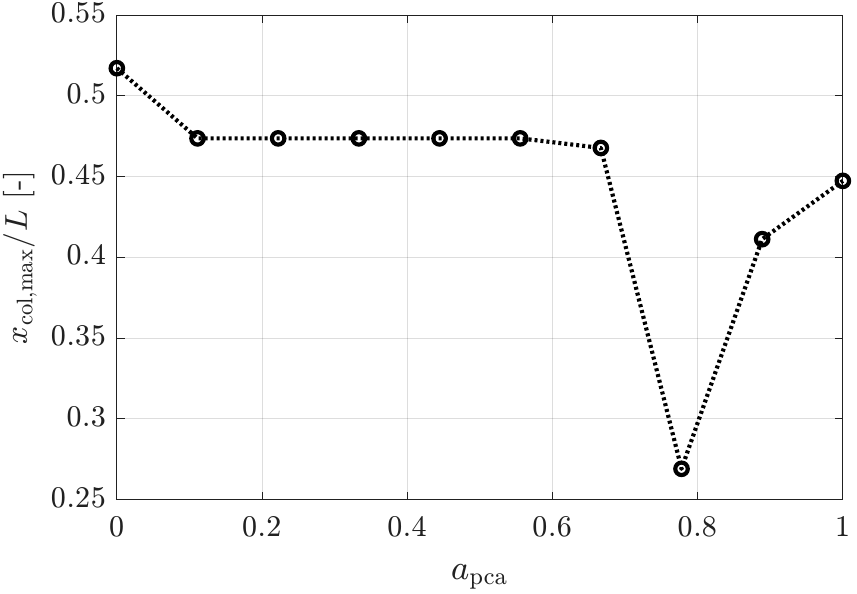} 
    \caption{
    \textbf{Influence of PCA muscle activation on vocal fold collision dynamics.}
Left: cycle-averaged maximum contact pressure $p_{\mathrm{col,max}}$. Right: corresponding normalized anterior--posterior location of the maximum contact pressure.
}
    \label{fig:PGOAnalysis} 
\end{figure}

Figure~\ref{fig:PGOAnalysis} illustrates the resulting variation in the maximum contact pressure and its anterior--posterior location. The results show that increasing PCA activation initially produces a modest increase in the maximum contact pressure, followed by a slight decrease. However, beyond a critical activation level, the maximum contact pressure increases substantially, suggesting that VF collision dynamics are a nonlinear function of posterior glottal opening.

The corresponding contact location also varies non-monotonically with PCA activation. For low to moderate activation levels, the location remains nearly unchanged near the anterior--posterior midpoint. Near $a_{\mathrm{pca}}\approx0.7$, however, the peak contact location undergoes a pronounced anterior shift, followed by a cusp-like variation near $a_{\mathrm{pca}}\approx0.8$ and a subsequent posterior migration at higher activation levels. This behavior suggests a shift in the underlying vibration patterns and collision mechanics as the posterior glottal gap increases.

\begin{figure}[ht!]
    \centering
        $$(a_{\mathrm{lca}},
        a_{\mathrm{ia}},
        a_{\mathrm{pca}}, a_{\mathrm{ct}},
        a_{\mathrm{ta}},
        )=(0.8,0.8,1,0.3,0.4)$$
    \includegraphics[width=0.425\linewidth]{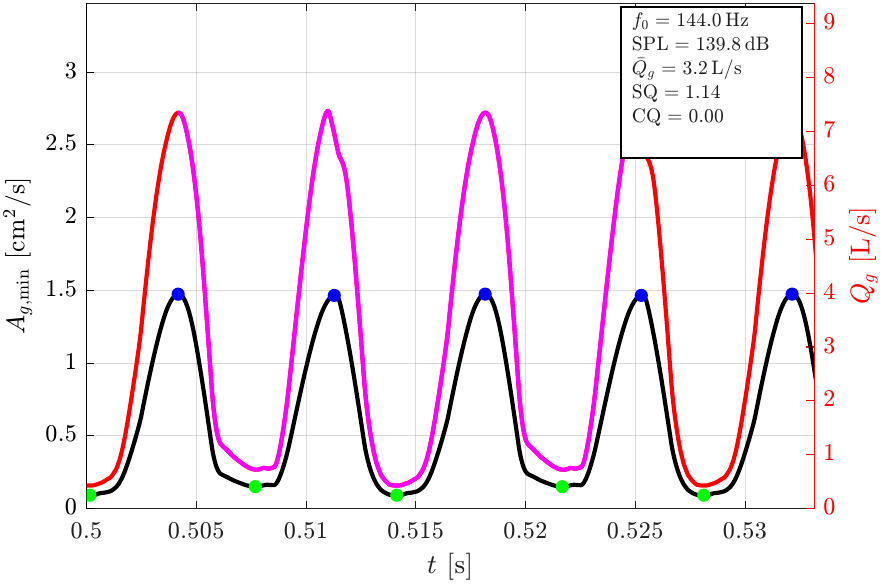}~\includegraphics[width=0.4\linewidth]{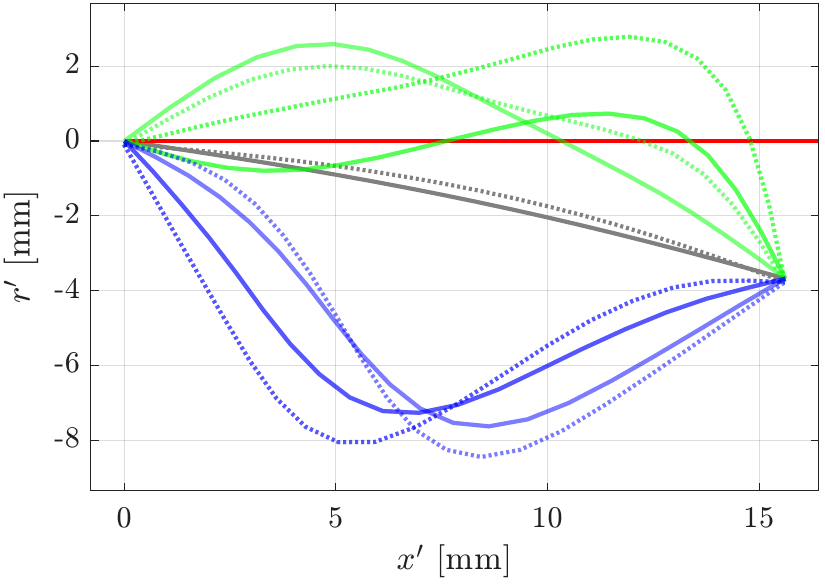}
    \includegraphics[width=0.8\linewidth]{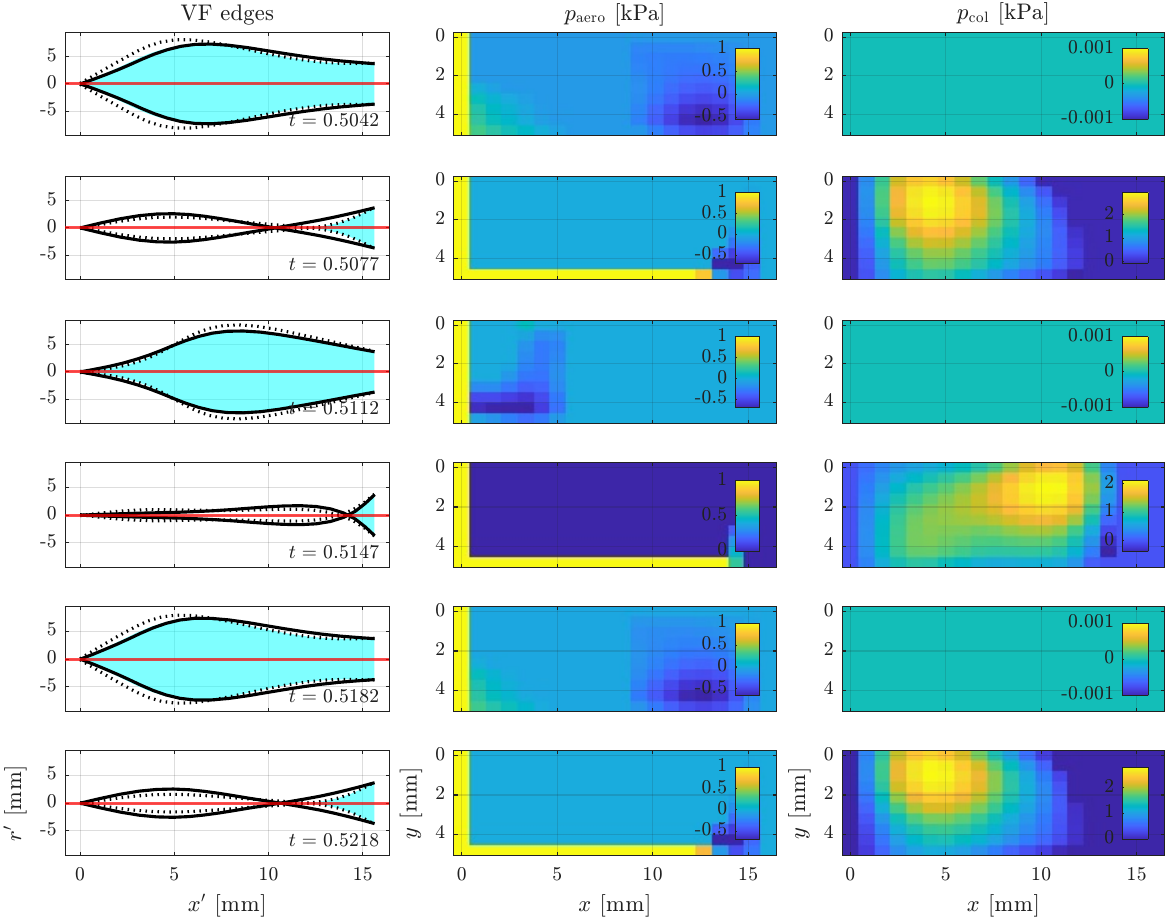}

    \caption{
    \textbf{Sustained phonation with higher order modes.}
    Top-left: glottal area and glottal flow-rate waveforms. The magenta-highlighted segment identifies the oscillation cycles illustrated in the lower panels. Top-right: left vocal fold medial-surface profiles corresponding to maximum opening (blue), maximum closure (green), and the static equilibrium configuration (gray). The maximum-opening and maximum-closure profiles correspond to the highlighted peak and nadir of the glottal area waveform, respectively. Solid and dashed curves denote the superior and inferior vocal fold edges, respectively, while the red line indicates the medial plane. Bottom: evolution of the vocal fold and glottal configuration, aerodynamic pressure distribution, and contact-pressure distribution over one representative oscillation cycle. The first column of panels shows the medial-surface shapes of both vocal folds, with the glottal opening shaded in cyan; gaps not shaded in cyan are completely closed (the folds are overlapping). The second and third columns show the corresponding aerodynamic and contact-pressure distributions, respectively.}

    \label{fig:Example3}
\end{figure}

To gain further insight into the effects of posterior glottal opening on phonation dynamics, we examine the case $a_{\mathrm{
pca}}=1$, which corresponds to the peak contact pressure predicted in Fig.~\ref{fig:PGOAnalysis} (left panel).  The  phonation dynamics for $a_{\mathrm{pca}}=1$ are shown in Fig.~\ref{fig:Example3}. Although the glottal area and glottal flow-rate waveforms remain periodic and qualitatively similar to those of Fig.~\ref{fig:Example2}, the underlying VF kinematics differ substantially. In particular, the medial-surface profiles exhibit multiple distinct configurations during glottal opening and closure. Inspection of the medial-surface evolution over successive oscillation cycles reveals the emergence of higher-order anterior--posterior vibration modes that are not present in the moderate-PCA case. Correspondingly, the contact pressure distribution exhibits pronounced anterior--posterior variations, with the location of peak contact alternating between different regions of the vocal folds throughout the cycle.  We note that this alternating pattern is also present for $a_{\mathrm{pca}}=0.8$, where Fig.~\ref{fig:PGOAnalysis} reveals pronounced changes in the contact-pressure behavior.

Overall, the results demonstrate that posterior glottal opening induced by PCA activation alters not only the magnitude of VF collision forces, but also the underlying vibration patterns that govern their spatial distribution and temporal dynamics. These findings highlight the intricate and strongly nonlinear interplay between glottal geometry, VF vibration, and collision mechanics during phonation.

\subsection*{Model performance, limitations, and future work}
We have presented a biomechanical VF model that establishes a direct relationship between laryngeal muscle activation, VF geometry/glottal conformation, and tissue mechanical properties, with sufficiently modest computational expense for broad-scale phonation simulations. The proposed framework predicts static VF configurations that are qualitatively consistent with those reported in previous numerical and clinical studies. Furthermore, the phonatory measures generated by the model exhibit trends that are in good qualitative agreement with those obtained from high-fidelity computational models. Through a representative case study, we demonstrate the capacity of the framework to explore clinically relevant problems, such as the influence of posterior glottal opening on phonatory biomechanics, which highlights the important role of glottal geometry in shaping collision dynamics and overall phonatory behavior.

Despite these encouraging results, there are several underlying assumptions and limitations worth noting. First, the model at present does not incorporate two-way acoustic coupling between the glottal source and the vocal tract. Consequently, the range of fundamental frequencies predicted by the model does not extend to the higher values reported in clinical studies and some numerical simulations \cite{AlzamendiPetersonErathHillmanZanartu21,martinez2025toward}. Second, the model does not reproduce the initial increase in $f_o$ with increasing TA activation at low TA activation levels that has been observed in previous computational and clinical studies \cite{ChhetriNeubauerSoferBerry14,MovahhediGengXueZheng21,AlzamendiManriquezHadwinDengPetersonErathMehtaHillmanZanartu20}. These discrepancies suggest that additional refinement of both the model structure and parameterization may be necessary to fully capture the complex interplay between laryngeal muscle activation and phonatory dynamics. Nevertheless, the framework captures many of the essential biomechanical mechanisms underlying phonation and reproduces a number of physiologically relevant qualitative trends. Combined with its computational efficiency and strong biomechanical interpretability, these characteristics make the proposed model a promising tool for future investigations of voice production.

Future work will focus on several extensions of the present framework. In particular, we aim to incorporate two-way coupling between the proposed VF model and the posturing framework of \cite{TitzeHunter07}, thereby enabling dynamic interactions between laryngeal muscle activation, posturing dynamics, and phonatory behavior. Additional developments will include refinement of the boundary conditions, constitutive assumptions, and contact and aerodynamic models to further improve physiological realism and predictive accuracy. We also aim to incoporporate extrinsic muscle effects in future iterations \cite{serry2023modeling}.

The proposed framework will ultimately be employed to investigate voice disorders associated with abnormal muscle activation patterns and inefficient vocal behaviors, including vocal hyperfunction \cite{HillmanSteppVanStanZanartuMehta20}. The computational efficiency of the model will enable extensive parametric studies to systematically examine the influence of muscle activation patterns, tissue properties, and glottal configurations on phonatory outcomes without sacrificing spatial information relevant to certain voice disorders.

\section*{Materials and methods}
\subsection*{Model development}
Throughout this section, quantities associated with the resting (unstrained) VF configuration are denoted by the subscript $(\cdot)_0$, whereas quantities without subscripts correspond to the prestrained configuration introduced in Model Overview.

For compactness, we introduce the index sets
\begin{equation}
\mathcal{I}
=
\{\mathrm{muc},\mathrm{lig},\mathrm{ta}\},
\qquad
\mathcal{J}
=
\{\mathrm{m},\mathrm{b}\},
\qquad
\mathcal{I}_{\mathrm{b}}
=
\{\mathrm{lig},\mathrm{ta}\},
\qquad
\mathcal{I}_{\mathrm{m}}
=
\{\mathrm{muc}\},
\end{equation}
where $\mathrm{muc}$, $\mathrm{lig}$, and $\mathrm{ta}$ denote the mucosa, vocal ligament, and thyroarytenoid muscle, respectively, while $\mathrm{m}$ and $\mathrm{b}$ denote the membrane and beam components of the model.

Let $L_0$ denote the resting VF length. The longitudinal stretch ratio is related to the nominal strain $\bar{\varepsilon}$ through the relation:
\begin{equation}
\label{eq:StretchRatio}
\lambda:=1+\bar{\varepsilon}.
\end{equation}
The corresponding VF length in the prestrained configuration is therefore
\begin{equation}
\label{eq:VFElongation}
L=\lambda L_0.
\end{equation}

The stretch ratio $\lambda$ is a fundamental parameter in the proposed framework, as it characterizes the prestressed VF configuration and governs the transformation between the resting and prestrained states. Consequently, $\lambda$ appears throughout the model development, introducing scaling factors in the geometric, kinematic, constitutive, and inertial quantities appearing in the governing equations.

Let $b_0$ denote the resting VF thickness. Each layer is assumed to possess a rectangular cross-section with resting area $A_{i,0}$ and thickness $b_{i,0}$. For simplicity, all layers are assumed to have the same resting thickness, namely
\[
b_{i,0}=b_0,
\qquad
i\in\mathcal{I}.
\]
The corresponding resting depth of each layer is then given by
\[
d_{i,0}
=
\frac{A_{i,0}}{b_{i,0}},
\qquad
i\in\mathcal{I}.
\]

Assuming incompressibility, longitudinal elongation is accompanied by a transverse contraction, yielding the prestrained dimensions
\begin{equation}
\label{eq:PrestrainedDimensions}
d_i
=
\frac{d_{i,0}}{\sqrt{\lambda}},
\qquad
b_i
=
\frac{b_{i,0}}{\sqrt{\lambda}}
=
b,
\qquad
A_i
=
\frac{A_{i,0}}{\lambda},
\qquad
i\in\mathcal{I}.
\end{equation}

The density of layer $i$ is denoted by $\rho_i$. Under the incompressibility assumption, the density remains unchanged by the prestrain. The resting geometric parameters and densities adopted throughout this work are summarized in Table~\ref{tab:RestingDimensions}.

\begin{table}[ht!]
\centering
\caption{\textbf{Resting dimensions and densities.
The cross-sectional areas are taken from \cite{TitzeAlipour06}. The resting length corresponds to a male vocal fold and falls within the physiological range reported for adult males \cite{su2002measurement}. The resting width $b_{0}$ is chosen as a representative value within the range of effective vocal-fold thicknesses measured in the absence of TA activation \cite{lehoux2024methodology}. The layer densities are adapted from \cite{StoryTitze95} and adjusted to preserve the physiologically plausible ordering $\rho_{\mathrm{muc}} < \rho_{\mathrm{lig}} < \rho_{\mathrm{ta}}$.} }
\begin{tabular}{ccc}
\hline
Parameter & Value & Units \\
\hline
$L_{0}$ & 1.50$\times10^{-2}$ & m \\
$b_{0}$ & 5.0$\times10^{-3}$ & m \\
$A_{\mathrm{muc},0}$ & 5.0$\times10^{-6}$ & m$^2$ \\
$A_{\mathrm{lig},0}$ & 6.10$\times10^{-6}$ & m$^2$ \\
$A_{\mathrm{ta},0}$ & 4.09$\times10^{-5}$ & m$^2$ \\
$\rho_{\mathrm{muc}}$ & 1.0$\times10^{3}$ & kg/m$^3$ \\
$\rho_{\mathrm{lig}}$ & 1.03$\times10^{3}$ & kg/m$^3$ \\
$\rho_{\mathrm{ta}}$ & 1.05$\times10^{3}$ & kg/m$^3$ \\
$\theta_{\mathrm{conv}}$ & 2.0$\times10^{-2}$ & rad \\
\hline
\end{tabular}
\label{tab:RestingDimensions}
\end{table}

Let $(x_{0},y_{0},r_{0})$ denote material coordinates in the resting configuration, where
\begin{itemize}
    \item $x_{0}\in[0,L_0]$ is the longitudinal coordinate,
    \item $y_{0}\in[0,b_0]$ is the inferior--superior coordinate,
    \item $r_{0}$ is the depth coordinate measured from the base of the TA muscle.
\end{itemize}

Under the imposed longitudinal prestretch, the VF deforms from the resting configuration into the prestrained configuration. The corresponding spatial coordinates $(x,y,r)$, illustrated in Figure~\ref{fig:SchematicMembraneBeam}, are related to the material coordinates through
\begin{equation}
\label{eq:CoordinateTransformation}
x=\lambda x_{0},
\qquad
y=\frac{1}{\sqrt{\lambda}}y_{0},
\qquad
r=\frac{1}{\sqrt{\lambda}}r_{0}.
\end{equation}

The coordinate transformation \eqref{eq:CoordinateTransformation}  induces corresponding transformations of spatial derivatives between the resting and prestrained configurations. In particular, application of the chain rule yields, for a function $f$,
\begin{equation}\label{eq:Scaling}
\partial_{x_0}f
=
\lambda\,\partial_x f,
\qquad
\partial_{y_0}f
=
\frac{1}{\sqrt{\lambda}}\partial_y f,
\qquad
\partial_{r_0}f
=
\frac{1}{\sqrt{\lambda}}\partial_r f,
\end{equation}
where $\partial_{\eta} f$ denotes the partial derivative of the function $f$ with respect to the variable $\eta$. Throughout this work, except when referring to the global coordinate system, primes denote differentiation with respect to the longitudinal coordinate $x$. Moreover, overdots denote differentiation with respect to time.
\medskip

\subsubsection*{Longitudinal strain fields }
 For each layer \(i\in\mathcal{I}\), let \(z_i\) denote the local transverse coordinate (along the \(r\)-axis) measured from the geometric center of the layer (see Fig.~\ref{fig:SchematicMembraneBeam}). The coordinate satisfies
\[
z_i \in \left[-\frac{d_i}{2},\,\frac{d_i}{2}\right],
\qquad
i\in\mathcal{I}.
\]

Let \(u_i\) denote the longitudinal displacement field of layer \(i\), measured relative to the prestrained configuration corresponding to the nominal strain \(\bar{\varepsilon}\). Furthermore, let \(\bar{u}_i\) denote the longitudinal displacement of the centroidal axis of layer \(i\), corresponding to \(z_i=0\).

Assuming a linear variation of the longitudinal displacement across the layer depth, the displacement field is approximated by
\begin{equation}
\label{eq:DisplacementField}
u_i
=
\bar{u}_i
-
z_i\,\psi_j,
\qquad
i\in\mathcal{I}_j,
\quad
j\in\mathcal{J},
\end{equation}
where \(\psi_j\) denotes the rotation variable associated with the kinematics of the corresponding structural component. Specifically, \(\psi_{\mathrm b}\) represents the rotation of a beam cross-sectional plane about the \(y\)-axis due to bending, whereas \(\psi_{\mathrm m}\) characterizes the rotation of a material line initially normal to the membrane midsurface in the longitudinal--transverse (\(x\)--\(r\)) plane. Positive values of \(\psi_{\mathrm b}\) and \(\psi_{\mathrm m}\) correspond to counterclockwise rotations in the \(-y\) direction. The precise relationship between these rotation variables and the displacement field will be established through the shear-deformation assumptions introduced later. In particular, the beam component is assumed to satisfy the Euler--Bernoulli hypothesis of vanishing transverse shear deformation, allowing \(\psi_{\mathrm b}\) to be expressed directly in terms of the transverse displacement. In contrast, transverse shear deformation is retained in the membrane component, and consequently \(\psi_{\mathrm m}\) remains an independent kinematic variable.

The membrane kinematic variables, \(\bar{u}_{\mathrm{muc}}\) and \(\psi_{\mathrm{m}}\), depend on the longitudinal coordinate \(x\), transverse coordinate \(y\), and time \(t\). In contrast, the beam variables, \(\bar{u}_{\mathrm{lig}}\), \(\bar{u}_{\mathrm{ta}}\), and \(\psi_{\mathrm{b}}\), depend only on \(x\) and \(t\). 

Continuity of the longitudinal displacement field at the ligament--TA interface imposes the kinematic compatibility condition
\begin{equation}\label{eq:ContinuityCondition}
\bar{u}_{\mathrm{ta}}
=
\bar{u}_{\mathrm{lig}}
+
\frac{1}{2}
\left(
d_{\mathrm{lig}}
+
d_{\mathrm{ta}}
\right)
\psi_{\mathrm{b}}.
\end{equation}
The longitudinal strain in layer $i$ is modeled using the engineering strain measured with respect to the resting configuration. Since the displacement field $u_i$ is defined relative to the prestrained configuration associated with the nominal strain $\bar{\varepsilon}$, the total longitudinal strain is given by
\begin{equation}\label{eq:GeneralStrainEquation}
\varepsilon_i
=
\bar{\varepsilon}
+
\lambda\,u'_i,
\qquad i\in\mathcal{I}.
\end{equation}

Substituting \eqref{eq:DisplacementField} into \eqref{eq:GeneralStrainEquation} yields
\begin{equation}\label{eq:StrainField}
\begin{aligned}
\varepsilon_i
&=
\bar{\varepsilon}
+
\lambda
\left(
\bar{u}'_i
-
z_i\,\psi_{j}'
\right),
\qquad i\in\mathcal{I}_{j},
~j\in \mathcal{J}.
\end{aligned}
\end{equation}
\subsubsection*{Longitudinal stress fields}

Since the equilibrium equations are formulated with respect to the prestrained configuration, the internal forces and moments must be expressed in terms of stresses defined per unit area of that configuration.

The material constitutive relations, however, are specified through engineering stress--strain laws referenced to the resting configuration. Consequently, it is necessary to construct a stress measure that is compatible with the prestrained configuration while remaining consistent with the underlying constitutive description.

Let
\[
\sigma_i^{\mathrm{e}}
=
\bar{\sigma}_i(\varepsilon_i)
\]
denote the longitudinal engineering stress of layer \(i\), where \(\bar{\sigma}_i\) is the corresponding nonlinear constitutive law and \(\sigma_i^{\mathrm{e}}\) represents force per unit area in the resting configuration.

To obtain a stress measure suitable for equilibrium calculations in the prestrained configuration, we invoke force equivalence,
\[
\sigma_i\,\mathrm{d}A
=
\sigma_i^{\mathrm{e}}\,\mathrm{d}A_0,
\]
where \(\mathrm{d}A_0\) and \(\mathrm{d}A\) denote infinitesimal cross-sectional area elements in the resting and prestrained configurations, respectively. Under the incompressibility assumption,
\[
\mathrm{d}A
=
\frac{\mathrm{d}A_0}{\lambda},
\]
which yields
\begin{equation}
\label{eq:PrestrainedStress}
\sigma_i
=
\lambda\,\sigma_i^{\mathrm{e}}.
\end{equation}

The quantity \(\sigma_i\) should therefore be interpreted as an effective stress measure referenced to the prestrained configuration rather than the exact Cauchy stress. This approximation is consistent with the small-on-large kinematic framework adopted herein, wherein the dynamic strains associated with phonation are assumed to remain infinitesimal relative to the nominal prestrain \(\bar{\varepsilon}\). As a result, the additional changes in cross-sectional area induced by the membrane and beam displacements are negligible, allowing the cross-sectional areas of the prestrained configuration to be treated as effectively constant in the evaluation of stress resultants and internal forces.

The engineering constitutive relation is linearized about the nominal strain $\bar{\varepsilon}$:
\begin{equation}\label{eq:LinearizedStress}
\sigma_i^{\mathrm{e}}
\approx
\bar{\sigma}_i(\bar{\varepsilon})
+
E_i\,(\varepsilon_i-\bar{\varepsilon}),
\qquad i\in\mathcal{I},
\end{equation}
where
\[
E_i
=
\frac{\mathrm{d}\bar{\sigma}_i}{\mathrm{d}\varepsilon}
\biggr\rvert_{\varepsilon=\bar{\varepsilon}},
\qquad
i\in\{\mathrm{muc},\mathrm{lig}\},~
E_{\mathrm{ta}}
=
\frac{\partial \bar{\sigma}_{\mathrm{ta}}}{\partial \varepsilon}
\biggr\rvert_{\varepsilon=\bar{\varepsilon},\,\texttt{a}=\texttt{a}_{\mathrm{ta}}}.
\]

Combining \eqref{eq:StrainField}, \eqref{eq:PrestrainedStress}, and \eqref{eq:LinearizedStress},  we obtain
\begin{equation}\label{eq:FinalTrueStressLinearApproximation}
\sigma_i
\approx
\lambda\bar{\sigma}_i(\bar{\varepsilon})
+
\lambda^2
E_i
\bigl(
\bar{u}_i'
-
z_i \psi_{j}'
\bigr),
\qquad
i\in\mathcal{I}_{j},
~j\in \mathcal{J}.
\end{equation}

Following \cite{AlzamendiPetersonErathHillmanZanartu21,HunterTitze07}, the engineering stress--strain constitutive relations are of exponential type, which are defined as follows. For each layer $i\in\mathcal{I}$, the passive engineering stress is modeled by
\begin{equation}
\bar{\sigma}_{p,i}(\varepsilon)
=
\begin{cases}
\displaystyle
-\frac{\sigma_{1,i}}{\varepsilon_{1,i}}
\left(
\varepsilon-\varepsilon_{1,i}
\right),
&
\varepsilon \leq \varepsilon_{2,i},
\\[10pt]
\displaystyle
-\frac{\sigma_{1,i}}{\varepsilon_{1,i}}
\left(
\varepsilon-\varepsilon_{1,i}
\right)
\\[4pt]
\displaystyle\qquad
+
\sigma_{2,i}
\Big(
\exp\big(
\alpha_{1,i}(\varepsilon-\varepsilon_{2,i})
\big)
-
1
-
\alpha_{1,i}(\varepsilon-\varepsilon_{2,i})
\Big),
&
\varepsilon \geq \varepsilon_{2,i}.
\end{cases}
\end{equation}
Active stress is assumed to be present only in the TA muscle and is modeled as \cite{TitzeAlipour06}
\begin{equation}
\bar{\sigma}_{a,\mathrm{ta}}(\varepsilon)
=
a_{\mathrm{ta}}
\,\sigma_{a,\max,\mathrm{ta}}
\,
\max
\left\{
0,
~
1
-
\alpha_{2,\mathrm{ta}}
(\varepsilon-\varepsilon_{m,\mathrm{ta}})^2
\right\}.
\end{equation}
Here, $\sigma_{1,i}$, $\sigma_{2,i}$, $\varepsilon_{1,i}$, $\varepsilon_{2,i}$, $\varepsilon_{m,i}$, $\alpha_{1,i}$, and $\alpha_{2,i}$ denote material parameters whose adopted numerical values are provided in Table~\ref{tab:MechanicalProperties}. These values are adapted from \cite{AlzamendiPetersonErathHillmanZanartu21} and adjusted to yield more physiologically realistic biomechanical behavior.
\begin{table}[h!]
\centering
\caption{\textbf{Constitutive relations parameters adapted from \cite{AlzamendiPetersonErathHillmanZanartu21}}.}
\begin{tabular}{ccc}
\hline
Parameter & Value & Units \\
\hline
$\sigma_{1,\mathrm{muc}}$ & 1.0$\times10^{3}$ & Pa \\
$\sigma_{1,\mathrm{lig}}$ & 5.0$\times10^{2}$ & Pa \\
$\sigma_{1,\mathrm{ta}}$ & 2.0$\times10^{3}$ & Pa \\
$\varepsilon_{1,\mathrm{muc}}$ & -5.0$\times10^{-1}$ & - \\
$\varepsilon_{1,\mathrm{lig}}$ & -5.0$\times10^{-1}$ & - \\
$\varepsilon_{1,\mathrm{ta}}$ & -5.0$\times10^{-1}$ & - \\
$\sigma_{2,\mathrm{muc}}$ & 1.0$\times10^{3}$ & Pa \\
$\sigma_{2,\mathrm{lig}}$ & 1.0$\times10^{3}$ & Pa \\
$\sigma_{2,\mathrm{ta}}$ & 1.0$\times10^{3}$ & Pa \\
$\varepsilon_{2,\mathrm{muc}}$ & -3.0$\times10^{-1}$ & - \\
$\varepsilon_{2,\mathrm{lig}}$ & -3.0$\times10^{-1}$ & - \\
$\varepsilon_{2,\mathrm{ta}}$ & -5.0$\times10^{-2}$ & - \\
$\alpha_{1,\mathrm{muc}}$ & 6.0 & - \\
$\alpha_{1,\mathrm{lig}}$ & 7.0 & - \\
$\alpha_{1,\mathrm{ta}}$ & 9.0 & - \\
$\sigma_{a,\max,\mathrm{ta}}$ & 1.0$\times10^{5}$ & Pa \\
$\varepsilon_{m,\mathrm{ta}}$ & 3.0$\times10^{-1}$ & - \\
$\alpha_{2,\mathrm{ta}}$ & 1.0 & - \\
\hline
\end{tabular}
\label{tab:MechanicalProperties}
\end{table}

The total engineering stress in each layer is therefore given by
\begin{equation}
\bar{\sigma}_i
=
\begin{cases}
\bar{\sigma}_{p,i},
&
i\in\{\mathrm{muc},\mathrm{lig}\},
\\[4pt]
\bar{\sigma}_{p,\mathrm{ta}}
+
\bar{\sigma}_{a,\mathrm{ta}},
&
i=\mathrm{ta}.
\end{cases}
\end{equation}
\subsubsection*{Internal forces and moments }
We now derive the axial forces and bending moments in each layer.

The axial force in layer $i$ is defined by
$N_i
=
\int_{A_i}
\sigma_i\,\mathrm{d}A_{i}
=
b_i
\int_{-d_i/2}^{d_i/2}
\sigma_i\,\mathrm{d}z_i,~
i\in\mathcal{I}$. 
Substituting the stress approximation \eqref{eq:FinalTrueStressLinearApproximation} yields
\[
N_i
=
b_i
\int_{-d_i/2}^{d_i/2}
\left[
\lambda\bar{\sigma}_i(\bar{\varepsilon})
+
\lambda^2
E_i
\bigl(
\bar{u}_i'
-
z_i \psi_j'
\bigr)
\right]
\mathrm{d}z_i,
\qquad
i\in\mathcal{I}_j,
~j\in\mathcal{J}.
\]
Using the identities
$\int_{-d_i/2}^{d_i/2}
\mathrm{d}z_i
=
d_i$ and
$\int_{-d_i/2}^{d_i/2}
z_i\,\mathrm{d}z_i
=
0$,
we obtain the axial-force expression in the prestrained configuration:
\begin{equation}\label{eq:NormalForcesPrestrained}
N_i
=
\lambda A_i\,\bar{\sigma}_i(\bar{\varepsilon})
+
\lambda^2 E_i A_i\,\bar{u}_i',
\qquad
i\in\mathcal{I}.
\end{equation}
The first term represents the prestress-induced axial force associated with the nominal strain $\bar{\varepsilon}$, whereas the second term corresponds to the incremental elastic force generated by dynamic longitudinal deformation relative to the prestrained configuration. The total axial force in the beam component is therefore given by
\begin{equation}\label{eq:TotalNormalForce}
\begin{aligned}
N_{\mathrm b}
&=
\sum_{i\in\mathcal{I}_{\mathrm b}}
\lambda A_i\,\bar{\sigma}_i(\bar{\varepsilon})
+
\lambda^2 E_i A_i\,\bar{u}_i'.
\end{aligned}
\end{equation}

The longitudinal stresses generate bending about an axis parallel to ($y$) direction. The corresponding bending moment about the centroidal axis of layer $i$ is defined by  $M_i
=
-
\int_{A_i}
z_i\,\sigma_i\,\mathrm{d}A_i
=
-
b_i
\int_{-d_i/2}^{d_i/2}
z_i\,\sigma_i\,\mathrm{d}z_i$.
Substituting \eqref{eq:FinalTrueStressLinearApproximation} into the definition of $M_i$ yields
\[
M_i
=
-
b_i
\int_{-d_i/2}^{d_i/2}
z_i
\left[
\lambda \bar{\sigma}_i(\bar{\varepsilon})
+
\lambda^2
E_i
\bigl(
\bar{u}_i'
-
z_i \psi_{\mathrm b}'
\bigr)
\right]
\mathrm{d}z_i.
\]
Since $\int_{-d_i/2}^{d_i/2}
z_i\,\mathrm{d}z_i
=
0$, the prestress contribution and the centroidal axial-strain contribution do not generate bending moments about the layer centroid. Consequently,
\begin{equation}\label{eq:CentralMomentsPrestrained}
M_i
=
\lambda^2
E_i
I_i\,
\psi_{j}',
\qquad
i\in\mathcal{I}_{j},
~j\in \mathcal{J},
\end{equation}
where
$I_i
=
b_i
\int_{-d_i/2}^{d_i/2}
z_i^2\,\mathrm{d}z_i
=
{b_i d_i^3}/{12}$ denotes the second moment of area of layer $i$ in the prestrained configuration.

To obtain a membrane-type formulation for the mucosal layer, we assume that the membrane bending stiffness is negligible, namely
$E_{\mathrm{muc}}I_{\mathrm{muc}}
\approx 0$. Accordingly, we set
\[
M_{\mathrm{muc}}=0.
\]

In the following, we consider only the bending moments generated within the beam component. Furthermore, we assume that the axial forces within the membrane do not contribute to the beam moment balance.

Let $r_i$ denote the $r$-coordinate of the geometric center of layer $i$ in the prestrained configuration. Then
\[
r_{\mathrm{ta}}
=
\frac{d_{\mathrm{ta}}}{2},
\qquad
r_{\mathrm{lig}}
=
d_{\mathrm{ta}}
+
\frac{d_{\mathrm{lig}}}{2}.
\]

Moreover, let $r_{c,\mathrm b}$ denote an arbitrary reference point on the beam cross-section in the prestrained configuration. The total bending moment of the beam layers about $r_{c,\mathrm b}$ is then given by
\begin{equation}\label{eq:McPrestrained}
M_{c,\mathrm b}
=
\sum_{i\in\mathcal{I}_{\mathrm b}}
\left(
M_i
+
(r_{c,\mathrm b}-r_i)N_i
\right).
\end{equation}

Substituting \eqref{eq:CentralMomentsPrestrained} and \eqref{eq:NormalForcesPrestrained} into \eqref{eq:McPrestrained} yields
\begin{equation}\label{eq:McExpandedPrestrained}
\begin{aligned}
M_{c,\mathrm b}
&=
\sum_{i\in\mathcal{I}_{\mathrm b}}
\lambda^2
E_i
I_i\,\psi_{\mathrm b}'
\\
&\quad
+
\sum_{i\in\mathcal{I}_{\mathrm b}}
(r_{c,\mathrm b}-r_i)
\lambda^2
E_i
A_i\,\bar{u}_i'
\\
&\quad
+
\sum_{i\in\mathcal{I}_{\mathrm b}}
(r_{c,\mathrm b}-r_i)
\lambda
A_i\,
\bar{\sigma}_i(\bar{\varepsilon}).
\end{aligned}
\end{equation}

\subsubsection*{Membrane dynamic equations}

\begin{figure}[ht!]
    \centering
    \includegraphics[width=0.7\linewidth]{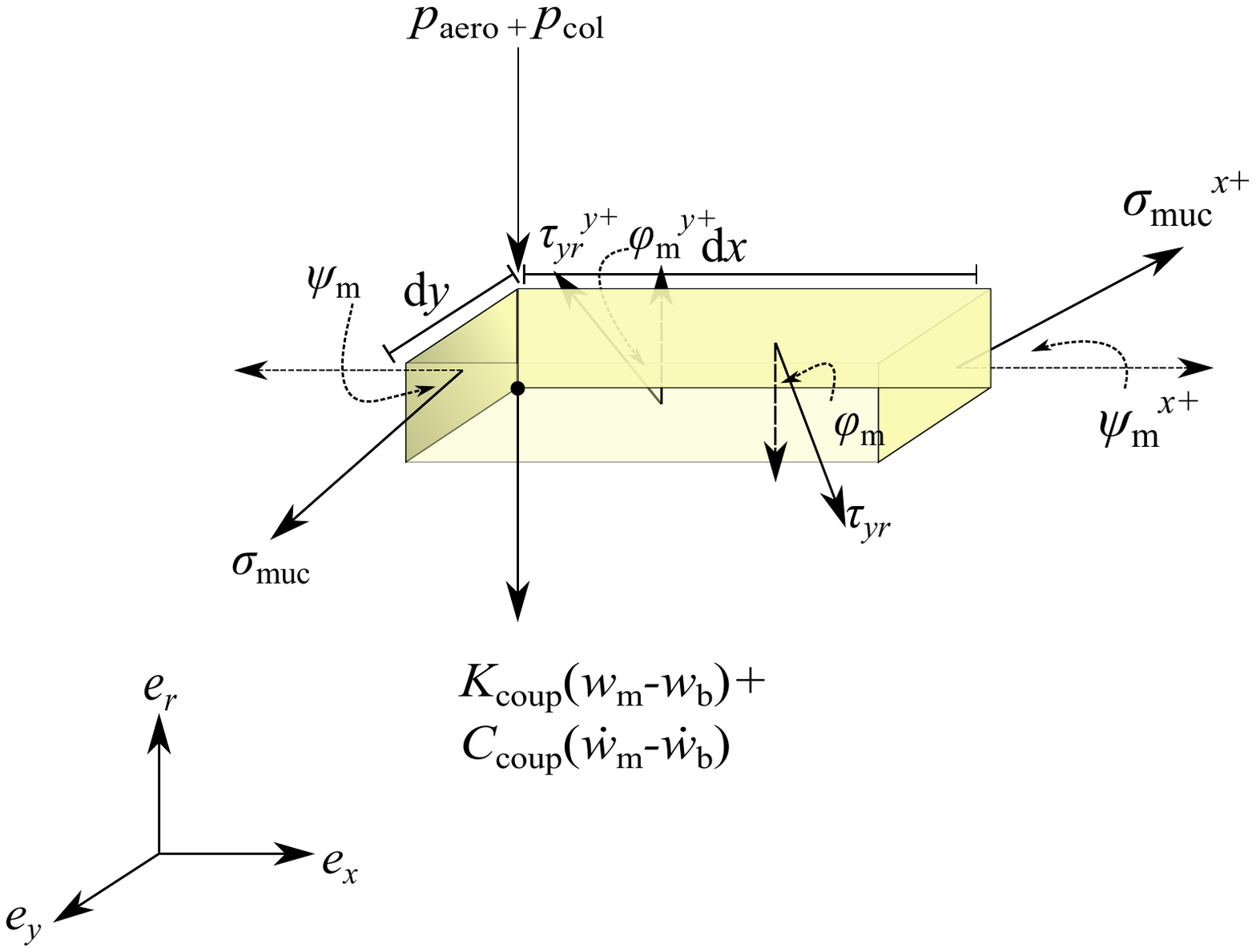}
    \caption{\textbf{Free-body diagram of an infinitesimal membrane element.}
Free-body diagram of an infinitesimal undeformed membrane element. The orientations of the normal and shear stresses are shown relative to the corresponding deformed configuration. The rotation angles $\psi_{\mathrm m}$ and $\varphi_{\mathrm m}$ characterize the local rotations associated with deformation in the longitudinal--transverse ($x$--$r$) and inferior--superior--transverse ($y$--$r$) planes, respectively. The superscripts $x+$ and $y+$ denote quantities evaluated at $(x+\mathrm{d}x,y,t)$ and $(x,y+\mathrm{d}y,t)$, respectively.}
    \label{fig:FBDMembraneNew}
\end{figure}

Consider an infinitesimal membrane element of dimensions $\mathrm{d}x\times \mathrm{d}y$ with lower-left corner located at $(x,y)$ (Figure~\ref{fig:FBDMembraneNew}). The membrane mass associated with this element is
$\mathrm{d}m_{\mathrm{m}}
=
\rho_{\mathrm{s,m}}\,
\mathrm{d}x\,\mathrm{d}y$, where
$\rho_{\mathrm{s,m}}
=
\rho_{\mathrm{muc}}\,d_{\mathrm{muc}}$
denotes the surface density of the mucosal layer in the prestrained configuration. Assuming small rotations, $\lvert \xi \rvert \ll 1,~
\xi\in\{\psi_{\mathrm m},\varphi_{\mathrm m}\}$,
we use the approximations
$\sin(\xi)\approx \xi$, 
$\cos(\xi)\approx 1$,  $\xi\in\{\psi_{\mathrm m},\varphi_{\mathrm m}\}$. Herein, $\psi_{\mathrm m}$ denotes the rotation angle of the membrane cross-section in the longitudinal--transverse ($x$--$r$) plane, while $\varphi_{\mathrm m}$ denotes the rotation angle associated with shear deformation in the inferior--superior--transverse ($y$--$r$) plane, corresponding to rotation about the longitudinal $x$-axis. For simplicity, we assume vanishing shear strain in the longitudinal--transverse plane, namely
\[
\gamma_{xr}
=
\partial_{x_0}w_{\mathrm m}
+
\partial_{z_0}u_{\mathrm{muc}}
=
\lambda \partial_x w_{\mathrm m}
-
\frac{\psi_{\mathrm m}}{\sqrt{\lambda}}
=
0.
\]
Hence, $$\psi_{\mathrm m}
=
\lambda^{3/2}
\partial_x w_{\mathrm m}.
$$

Let the superscripts $x+$ and $y+$ denote quantities evaluated at $(x+\mathrm{d}x,y,t)$ and $(x,y+\mathrm{d}y,t)$, respectively.

Neglecting in-plane inertia and external longitudinal loading, force balance in the longitudinal ($x$)-direction yields (see Figure~\ref{fig:FBDMembraneNew})
\begin{align*}
&
\sigma_{\mathrm{muc}}^{x+}
\cos(\psi_{\mathrm m}^{x+})
d_{\mathrm{muc}}\,\mathrm{d}y
-
\sigma_{\mathrm{muc}}
\cos(\psi_{\mathrm m})
d_{\mathrm{muc}}\,\mathrm{d}y\approx
d_{\mathrm{muc}}
\partial_x \sigma_{\mathrm{muc}}
\,
\mathrm{d}x\,\mathrm{d}y
=
0.
\end{align*} Consequently, $\partial_x \sigma_{\mathrm{muc}}
=
0$, so that $\sigma_{\mathrm{muc}}$ is independent of $x$. The net transverse contribution of the longitudinal stresses acting on the faces normal to the $x$-direction is
\begin{align*}
\sigma_{\mathrm{muc}}^{x+}
\sin(\psi_{\mathrm m}^{x+})
d_{\mathrm{muc}}\,\mathrm{d}y
-
\sigma_{\mathrm{muc}}
\sin(\psi_{\mathrm m})
d_{\mathrm{muc}}\,\mathrm{d}y
\approx
d_{\mathrm{muc}}
\lambda^{3/2}
\sigma_{\mathrm{muc}}
\partial_{xx}w_{\mathrm m}
\,
\mathrm{d}x\,\mathrm{d}y.
\end{align*}
To capture inferior--superior mucosal-wave propagation together with internal viscous dissipation, we introduce effective viscoelastic shear interactions between adjacent membrane elements in the inferior--superior ($y$) direction. The corresponding engineering shear strain, defined with respect to the resting configuration, is
\begin{equation}\label{eq:ShearStrainMembrane}\gamma_{yr}
=
\partial_{y_0}w_{\mathrm m}
+
\partial_{z_0}u_{\mathrm{muc}}
=
\frac{1}{\sqrt{\lambda}}
\left(
\partial_y w_{\mathrm m}
-
\varphi_{\mathrm m}
\right).
\end{equation}
We adopt the Kelvin--Voigt constitutive relation \cite{chan1999viscoelastic}
\begin{equation}\label{eq:KelvinModel}
\tau_{yr}
=
G_{\mathrm m}\gamma_{yr}
+
\eta_{\mathrm m}\dot{\gamma}_{yr},
\end{equation}
where $\tau_{yr}$ denotes the engineering shear stress acting in the transverse ($r$) direction on a face normal to the inferior--superior ($y$) direction, while $G_{\mathrm m}$ and $\eta_{\mathrm m}$ denote the elastic and viscous shear coefficients of the mucosal layer, respectively.

Since the engineering shear stress is defined per unit area in the resting configuration, the corresponding shear force acting on a face normal to the $y$-direction is $\mathrm{d}F_{\mathrm{shear}}
=
\tau_{yr}\,
\mathrm{d}x_0\,d_{\mathrm{muc},0}$. Using
$\mathrm{d}x_0
={\mathrm{d}x}/{\lambda}$ , 
$d_{\mathrm{muc},0}
=
\sqrt{\lambda}\,
d_{\mathrm{muc}}$, we obtain $\mathrm{d}F_{\mathrm{shear}}
=
\tau_{yr}\,
d_{\mathrm{muc}}\,
\mathrm{d}x/{\sqrt{\lambda}}.
$ Under the small-angle assumption, the shear forces act predominantly in the transverse ($r$)-direction. Taking the difference of the shear forces acting on opposite faces yields the net transverse shear force
${d_{\mathrm{muc}}}
\partial_y\tau_{yr}
\,
\mathrm{d}x\,\mathrm{d}y/\sqrt{\lambda}$.
Moreover, substituting \eqref{eq:ShearStrainMembrane} into \eqref{eq:KelvinModel},
\begin{equation}\label{eq:ShearStressMembrane}
\tau_{yr}
=
\frac{1}{\sqrt{\lambda}}
\left(
G_{\mathrm m}
(\partial_y w_{\mathrm m}-\varphi_{\mathrm m})
+
\eta_{\mathrm m}
(\partial_y\dot w_{\mathrm m}-\dot\varphi_{\mathrm m})
\right).
\end{equation}

Hence, the total transverse shear contribution becomes
\[
\left(
\frac{d_{\mathrm{muc}}}{\lambda}
\partial_y
\left(
G_{\mathrm m}
(\partial_y w_{\mathrm m}-\varphi_{\mathrm m})
+
\eta_{\mathrm m}
(\partial_y\dot w_{\mathrm m}-\dot\varphi_{\mathrm m})
\right)
\right)
\mathrm{d}x\,\mathrm{d}y.
\]
In addition, the membrane is subjected to aerodynamic pressure $p_{\mathrm{aero}}$, contact pressure $p_{\mathrm{col}}$, and beam--membrane interaction tractions of the form $K_{\mathrm{coup}}\bigl(w_{\mathrm m}-w_{\mathrm b}\bigr)$
and
$C_{\mathrm{coup}}\bigl(\dot w_{\mathrm m}-\dot w_{\mathrm b}\bigr)$, where $K_{\mathrm{coup}}$ denotes a spatially varying coupling stiffness in the inferior--superior direction, defined by
\[
K_{\mathrm{coup}}(y)=K_{\mathrm{coup},\max}+\bigl(K_{\mathrm{coup},\min}-K_{\mathrm{coup},\max}\bigr)\frac{y}{b},
\qquad y\in[0,b],
\]
with stiffness parameters satisfying $K_{\mathrm{coup},\min}<K_{\mathrm{coup},\max}$. This distribution yields a stiffer coupling near the superior edge and a softer coupling near the inferior edge, thereby promoting inferior-edge lead during the closing phase of vibration.  We note that similar, but lumped, versions of inferior-superior stiffness variations have been adopted in previous numerical phonation studies  \cite{JiangZhangStern01,TitzeStory02}. Moreover, $C_{\mathrm{coup}}\geq0$ is a viscous damping coefficient governing the dissipative beam--membrane interaction. Under the small-slope assumption, these loads act predominantly in the transverse ($r$)-direction. Applying Newton's second law in the transverse direction and dividing by $\mathrm{d}x\,\mathrm{d}y$ yields
\begin{equation}
\label{eq:MembraneEqnRefined}
\begin{split}
\rho_{\mathrm{s,m}}
\ddot{w}_{\mathrm m}
&=
\tilde{T}_{\mathrm{m}}
\partial_{xx}w_{\mathrm m}
\\
&\quad
+
\partial_y
\left(
\tilde{G}_{\mathrm m}
(\partial_y w_{\mathrm m}-\varphi_{\mathrm m})
+
\tilde{\eta}_{\mathrm m}
(\partial_y\dot w_{\mathrm m}-\dot\varphi_{\mathrm m})
\right)
\\
&\quad
-
K_{\mathrm{coup}}(w_{\mathrm m}-w_{\mathrm b})
-
C_{\mathrm{coup}}(\dot w_{\mathrm m}-\dot w_{\mathrm b})
\\
&\quad
-
p_{\mathrm{aero}}
-
p_{\mathrm{col}},
\end{split}
\end{equation}
where
$$
\tilde{T}_{\mathrm{m}}=d_{\mathrm{muc}}
\lambda^{3/2}
\sigma_{\mathrm{muc}},
~
\tilde{G}_{\mathrm{m}}=\frac{d_{\mathrm{muc}}}{\lambda}G_{\mathrm{m}},~
\tilde{\eta}_{\mathrm{m}}=\frac{d_{\mathrm{muc}}}{\lambda}\eta_{\mathrm{m}}.
$$

Throughout the derivation, higher-order terms in the membrane slopes and rotations have been neglected consistently under the small-angle assumption.

It remains to derive the dynamic equation governing the rotation angle $\varphi_{\mathrm m}$. Recall that this angle represents rotation about the longitudinal $x$-axis induced by shear deformation in the $y$--$r$ plane. The governing equation is obtained from angular momentum balance about the $x$-axis.

The mass moment of inertia of an infinitesimal membrane element about the $x$-axis is
$\mathrm{d}J_x
=
\rho_{\mathrm{muc}}\,
\mathrm{d}x\,\mathrm{d}y\,d_{\mathrm{muc}}
\frac{1}{12}
\left(
d_{\mathrm{muc}}^2+\mathrm{d}y^2
\right)$. Neglecting the higher-order contribution proportional to $\mathrm{d}y^2$ yields the approximation
$
\mathrm{d}J_x
\approx
\rho_{\mathrm{muc}}\,
\mathrm{d}x\,\mathrm{d}y\,d_{\mathrm{muc}}
{d_{\mathrm{muc}}^2}/{12}$. The shear stress $\tau_{yr}$ produces a restoring moment about the $x$-axis. Under the small-angle approximation, the corresponding distributed moment is approximately given by
$\mathrm{d}M_{\mathrm{shear}}=\mathrm{d}F_{\mathrm{shear}}\mathrm{d}y
=
{d_{\mathrm{muc}}}
\tau_{yr}\,
\mathrm{d}x\,\mathrm{d}y/{\sqrt{\lambda}}$. Therefore, angular momentum balance about the $x$-axis yields
\[
\rho_{\mathrm{muc}}\,
\mathrm{d}x\,\mathrm{d}y\,d_{\mathrm{muc}}
\frac{d_{\mathrm{muc}}^2}{12}
\ddot{\varphi}_{\mathrm m}
=
\frac{d_{\mathrm{muc}}}{\sqrt{\lambda}}
\tau_{yr}\,
\mathrm{d}x\,\mathrm{d}y.
\]
Dividing both sides by $\mathrm{d}x\,\mathrm{d}y\,d_{\mathrm{muc}}$, substituting   relation \eqref{eq:ShearStressMembrane}, and rearranging yield:
\begin{equation}\label{eq:RotationMembrane}
J_{\mathrm m}
\ddot{\varphi}_{\mathrm m}
=
G_{\mathrm m}
(\partial_y w_{\mathrm m}-\varphi_{\mathrm m})
+
\eta_{\mathrm m}
(\partial_y\dot w_{\mathrm m}-\dot\varphi_{\mathrm m}),
\end{equation}
where
\[
J_{\mathrm m}
=
\lambda
\rho_{\mathrm{muc}}
\frac{d_{\mathrm{muc}}^2}{12}.
\]
\subsubsection*{Beam dynamic equation}
\begin{figure}[ht!]
    \centering
    \includegraphics[width=0.6\linewidth]{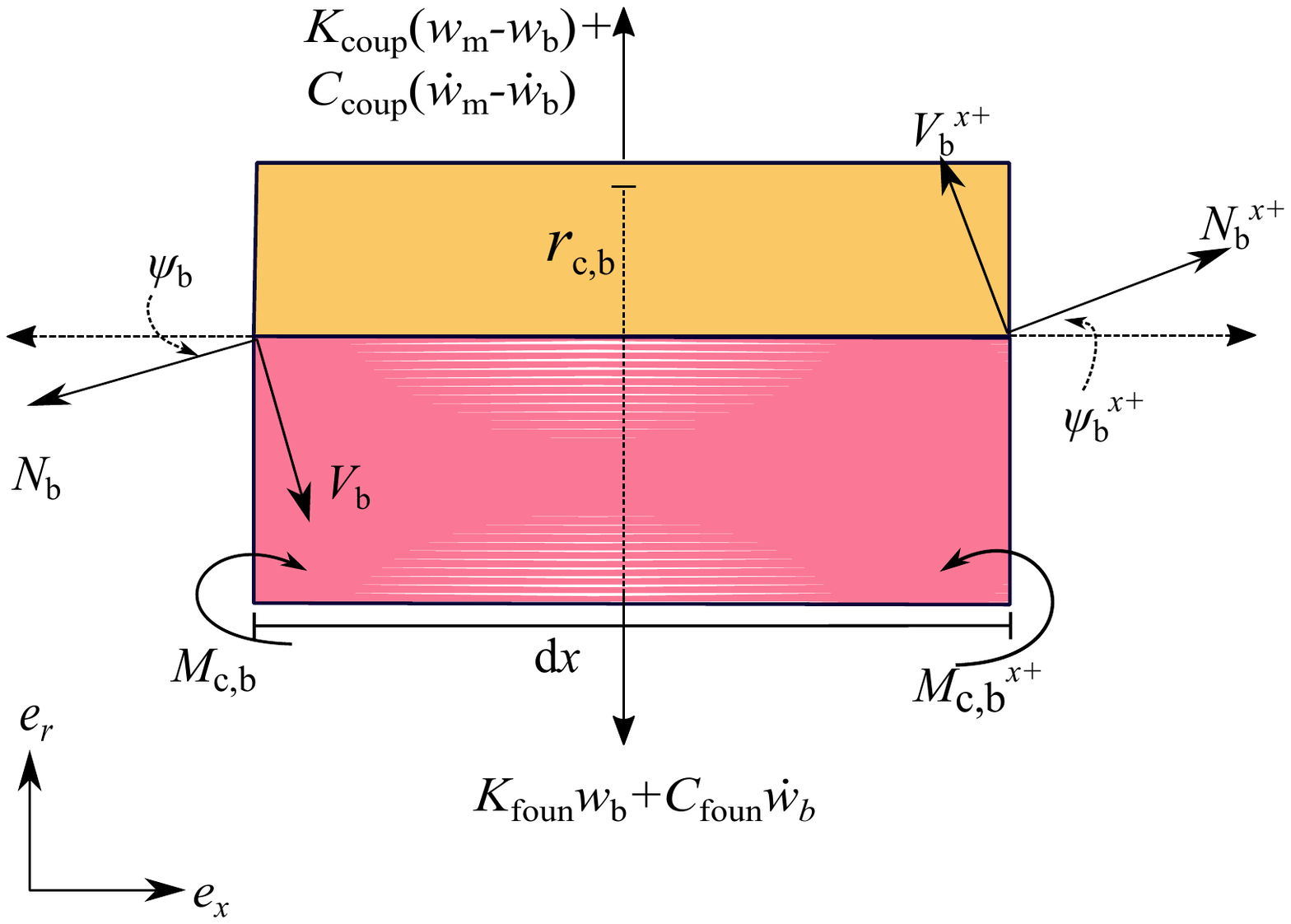}
    \caption{\textbf{Internal forces and moments acting on an infinitesimal beam element.}
Free-body diagram of an infinitesimal undeformed beam element. The directions of the internal forces and moments are shown relative to the corresponding deformed configuration, where $\psi_{\mathrm b}$ denotes the local cross-sectional rotation associated with beam bending. Superscripts such as $x+$ indicate quantities evaluated at $(x+\mathrm{d}x,t)$.}
    \label{fig:FBDBeamNew}
\end{figure}

Now, consider an infinitesimal beam element of length $\mathrm{d}x$ located at position $x$ (Figure~\ref{fig:FBDBeamNew}).

The beam mass is given by
$\mathrm{d}m_{\mathrm b}
=
\rho_{\mathrm{l,b}}\,
\mathrm{d}x$, where the line density is
$\rho_{\mathrm{l,b}}
=
\rho_{\mathrm{lig}}A_{\mathrm{lig}}
+
\rho_{\mathrm{ta}}A_{\mathrm{ta}}$.
 Let $V_{\mathrm b}(x,t)$ denote the internal shear force and recall that $M_{c,\mathrm b}(x,t)$ is the bending moment about the reference axis $r=r_{c,\mathrm b}$ (Eq. \eqref{eq:McExpandedPrestrained}). To obtain an Euler--Bernoulli-type beam model from the underlying shear-deformable kinematics, we impose rotational equilibrium,
\begin{equation}\label{eq:MomentBalanceBeam}
    \partial_x M_{c,\mathrm b}
+
V_{\mathrm b}
=
0,
\end{equation}
together with the Euler--Bernoulli constraint of vanishing engineering shear strain in the beam.

For $i\in\mathcal{I}_{\mathrm b}$, the engineering shear strain is defined with respect to the resting coordinates $(x_0,z_0)$. Using the prestrained coordinates $(x,z)$ together with the longitudinal stretch $\lambda$ and the scaling in  \eqref{eq:Scaling} , we obtain
\[
\gamma_{\mathrm b}
=
\partial_{x_0}w_{\mathrm b}
+
\partial_{r_0}u_i
=
\lambda \partial_x w_{\mathrm b}
+
\frac{1}{\sqrt{\lambda}}
\partial_r u_i
=
\lambda w_{\mathrm b}'
-
\frac{\psi_{\mathrm b}}{\sqrt{\lambda}}
=
0,
\]
which yields
\begin{equation}\label{eq:ZeroShearStrainBeam}
\psi_{\mathrm b}
=
\lambda^{3/2}
w_{\mathrm b}'.
\end{equation}

Since axial translation and longitudinal inertia are neglected, axial force equilibrium gives $\partial_x N_{\mathrm b}
=
0,
$
so that $N_{\mathrm b}$ is independent of $x$. Under a prescribed uniform nominal strain $\bar{\varepsilon}$, we therefore assume
\begin{equation}
\label{eq:ForceAssumptionRefined}
N_{\mathrm b}
=
\lambda
\sum_{i\in\mathcal{I}_{\mathrm b}}
A_i\,
\bar{\sigma}_i(\bar{\varepsilon}).
\end{equation}
By  comparing \eqref{eq:TotalNormalForce} with \eqref{eq:ForceAssumptionRefined}, we conclude that $\sum_{i\in\mathcal{I}_{\mathrm b}}
E_iA_i\,\bar{u}_i'
=
0$. Using this relation, in combination with the derivative of  the kinematic continuity condition \eqref{eq:ContinuityCondition}  gives
$\bar{u}_i'
=
\alpha_i\psi_{\mathrm b}',~
i\in\mathcal{I}_{\mathrm b}$,
where
\[
\alpha_{\mathrm{lig}}
=
-
\frac{
d_{\mathrm{ta}}+d_{\mathrm{lig}}
}{
2\left(
1+\dfrac{
E_{\mathrm{lig}}A_{\mathrm{lig}}
}{
E_{\mathrm{ta}}A_{\mathrm{ta}}
}
\right)
},
~
\alpha_{\mathrm{ta}}
=
\frac{
d_{\mathrm{ta}}+d_{\mathrm{lig}}
}{
2\left(
1+\dfrac{
E_{\mathrm{ta}}A_{\mathrm{ta}}
}{
E_{\mathrm{lig}}A_{\mathrm{lig}}
}
\right)
}.
\]
Substituting these expressions into the bending-moment relation  \eqref{eq:McExpandedPrestrained} yields
\begin{equation}
\label{eq:McSimplifiedRefined}
M_{c,\mathrm b}
=
\mu_{c,\mathrm b}\,
\psi_{\mathrm b}'
+
\bar{M}_{c,\mathrm b},
\end{equation}
where the effective bending stiffness is defined by
\[
\mu_{c,\mathrm b}
=
\lambda^2
\sum_{i\in\mathcal{I}_{\mathrm b}}
E_i
\left(
I_i
+
(r_{c,\mathrm b}-r_i)A_i\alpha_i
\right),
\]
and the nominal prestress-induced bending moment is
\[
\bar{M}_{c,\mathrm b}
=
\lambda
\sum_{i\in\mathcal{I}_{\mathrm b}}
(r_{c,\mathrm b}-r_i)
A_i\,
\bar{\sigma}_i(\bar{\varepsilon}).
\]

We now consider transverse force balance. The beam is subjected to:
(i) the geometric-stiffness contribution induced by the nominal axial force $N_{\mathrm b}$,
(ii) the gradient of the internal shear force,
(iii) a surrounding-tissue/foundation reaction modeled by $K_{\mathrm{foun}}w_{\mathrm b}+C_{\mathrm{foun}}\dot{w}_{\mathrm b}$, and
(iv) the beam--membrane interaction obtained by integrating the interface traction over $y\in[0,b]$.

Applying Newton's second law in the transverse direction yields
\begin{equation}
\begin{split}
\label{eq:LinearMomentumBeamRefined}
\rho_{\mathrm{l,b}}
\ddot{w}_{\mathrm b}
=&
N_{\mathrm b}\,
\partial_x\psi_{\mathrm b}
+
\partial_xV_{\mathrm b}
\\
&\quad
-
\int_0^b
\left[
K_{\mathrm{coup}}(w_{\mathrm b}-w_{\mathrm m})
+
C_{\mathrm{coup}}(\dot{w}_{\mathrm b}-\dot{w}_{\mathrm m})
\right]
\mathrm{d}y
\\
&\quad
-
K_{\mathrm{foun}}w_{\mathrm b}
-
C_{\mathrm{foun}}\dot{w}_{\mathrm b},
\end{split}
\end{equation}
where higher-order slope terms have been neglected consistently under the small-slope assumption. Using
$\partial_xV_{\mathrm b}
=
-
\partial_{xx}M_{c,\mathrm b}$ obtained by differentiating \eqref{eq:MomentBalanceBeam},
and substituting \eqref{eq:McSimplifiedRefined}  and  \eqref{eq:ZeroShearStrainBeam}, we get

\begin{equation}
\label{eq:BeamEqnRefined}
\begin{split}
\rho_{\mathrm{l,b}}
\ddot{w}_{\mathrm b}
=&
\tilde{N}_{\mathrm b}\,
\partial_{xx}w_{\mathrm b}
-
\tilde{\mu}_{c,\mathrm b}\,
\partial_{xxxx}w_{\mathrm b}
\\
&\quad
-
\int_0^b
\left[
K_{\mathrm{coup}}(w_{\mathrm b}-w_{\mathrm m})
+
C_{\mathrm{coup}}(\dot{w}_{\mathrm b}-\dot{w}_{\mathrm m})
\right]
\mathrm{d}y
\\
&\quad
-
K_{\mathrm{foun}}w_{\mathrm b}
-
C_{\mathrm{foun}}\dot{w}_{\mathrm b},
\end{split}
\end{equation}
where
\[
\tilde{N}_{\mathrm b}
=
\lambda^{3/2}N_{\mathrm b},~
\tilde{\mu}_{c,\mathrm b}
=
\lambda^{3/2}\mu_{c,\mathrm b}.
\]

Equations~\eqref{eq:MembraneEqnRefined}, \eqref{eq:RotationMembrane}, and \eqref{eq:BeamEqnRefined} constitute the coupled beam--membrane system under the stated assumptions.

\subsubsection*{Contact pressure}

Vocal fold contact is modeled using a unilateral viscoelastic foundation acting in the transverse ($r$) direction. In the present formulation, the beam--membrane system is described relative to the $x$--$y$ plane, which serves as the reference plane for the derivation. Consequently, the medial plane separating the left and right VFs must be expressed with respect to this coordinate system.

The medial plane is described by
\[
\mathbf{G}_{\mathrm{medial}}
=
x\tan(\theta_{g})
+
y\tan(\theta_{\mathrm{conv}}).
\]
Due to  assumed VF symmetry, contact occurs when the membrane displacement exceeds the medial-plane boundary, namely when $w_{\mathrm m}
>
\mathbf{G}_{\mathrm{medial}}$. 
The resulting contact pressure is modeled as
\begin{equation}
\label{eq:ContactModelRefined}
p_{\mathrm{col}}
=
\left(
K_{\mathrm{col}}
\bigl(
w_{\mathrm m}-\mathbf{G}_{\mathrm{medial}}
\bigr)
+
C_{\mathrm{col}}
\dot{w}_{\mathrm m}
\right)
\mathbf H
\bigl(
w_{\mathrm m}-\mathbf{G}_{\mathrm{medial}}
\bigr),
\end{equation}
where $K_{\mathrm{col}}>0$ and $C_{\mathrm{col}}>0$ denote the collision stiffness and damping coefficients, respectively, and $\mathbf H$ denotes the Heaviside function. Accordingly, the contact force is activated only when the membrane penetrates the medial plane.

\subsubsection*{Aerodynamic pressure}

 The glottal airflow is modeled using a quasi-steady viscous Bernoulli approximation (see, .e.g, \cite{VanDenBergZantemaDoornenbal57,PelorsonHirschbergVanHasselWijnandsAuregan94} ) with longitudinal sectioning along the $x$-direction. We assume a prescribed subglottal pressure $P_s$ and zero supraglottal pressure. For each fixed longitudinal position $x$, the flow is treated as one-dimensional in the inferior--superior direction  through a channel of width $\mathrm{d}x$. 

Assuming symmetry about the glottal midline together with small glottal and convergence angles, the local glottal gap height at $(x,y,t)$ is approximated by
\begin{equation}
\label{eq:LocalGapHeightRefined}
d_g
=
\max
\left\{
2\bigl(
\mathbf{G}_{\mathrm{medial}}-w_{\mathrm m}
\bigr),
\,0
\right\}.
\end{equation}
For a longitudinal slice of width $\mathrm{d}x$, the corresponding local cross-sectional flow area is
$\mathrm{d}A_g(x,y,t)
=
d_g(x,y,t)\,\mathrm{d}x $. For each fixed $x$, define the minimum gap height
\begin{equation}
\label{eq:MinimumGapRefined}
d_{\min}(x,t)
=
\min_{y\in[0,b]}
d_g(x,y,t).
\end{equation}
The local minimum glottal area is 
$
\mathrm{d}A_{g,\min}(x,t)
=
d_{\min}(x,t)\,\mathrm{d}x
$ 
and the minimum glottal area is 
$
A_{g,\min}=\int_{0}^{L}\mathrm{d}A_{g,\min}.
$

Assume initially that
 $d_{\min}>0$, corresponding to a locally open glottis. Let $y_{\min}$ denote a location where the minimum is attained; if multiple minimizers exist, the largest value of $y$ is selected.

Flow separation is assumed to occur when the local gap reaches a value proportional to the minimum gap height, namely
 $d_{\mathrm{sep}}
=
\alpha_s d_{\min}$,
where $\alpha_s>1$ is the separation ratio, which is in the range $\alpha_s\in[1.1,1.9]$ (see, e.g., \cite{alipour2004flow,PelorsonHirschbergVanHasselWijnandsAuregan94,SidlofDoareCadotChaigne11}). Let $y_{\mathrm{sep}}$ denote the corresponding separation location. The separation area for the slice is $\mathrm{d}A_{\mathrm{sep}}
=
d_{\mathrm{sep}}\,\mathrm{d}x $. For fixed $x$, let $\mathrm{d}q_g(x,t)$ denote the volume flow rate through the slice of width $\mathrm{d}x$. Since the coordinate system points in the upstream direction, the physical flow toward the superior edge corresponds to
$\mathrm{d}q_g<0.
$
By mass conservation, $\mathrm{d}q_g$ is independent of $y$ along the attached portion of the channel. Hence, $v
=
{\mathrm{d}q_g}/
{\mathrm{d}A_g}
=
{\mathrm{d}q_g}/
(d_g\,\mathrm{d}x)$. To account for viscous losses, we use the one-dimensional quasi-steady momentum balance (see, e.g., \cite{VanDenBergZantemaDoornenbal57,LuceroSchoentgen15})
\begin{equation}
\label{eq:MomentumBalanceFlow}
\frac{\partial p_{\mathrm{aero}}}{\partial y}
=
-\rho_a v\frac{\partial v}{\partial y}
-
\frac{12\mu_a\,\mathrm{d}q_g}
{\mathrm{d}x\,d_g^3},
\end{equation}
where $\rho_a$ is the air density and $\mu_a$ is the dynamic viscosity of air (see Table \ref{tab:AerodynamicParameters} for numerical values).
\begin{table}[h!]
\centering
\caption{\textbf{Aerodynamic parameters. Speed of sound, air density, and dynamic viscosity are adapted from \cite{LuceroSchoentgen15}. Separation ratio is tuned within reported ranges in the literature \cite{alipour2004flow}.}}
\begin{tabular}{ccc}
\hline
Parameter & Value & Units \\
\hline
$\alpha_{\mathrm{s}}$ & 1.30 & - \\
$\rho_{{a}}$ & 1.14 & kg/m$^3$ \\
$\mu_{{a}}$ & 1.80$\times10^{-5}$ & Pa.s \\
$c_{a}$ & 3.50$\times10^{2}$ & m/s \\
\hline
\end{tabular}
\label{tab:AerodynamicParameters}
\end{table}

The first term in the RHS of \eqref{eq:MomentumBalanceFlow} corresponds to the Bernoulli acceleration contribution, whereas the second term represents the Poiseuille/lubrication-type viscous pressure loss through a narrow slit. Using
$v\partial_{y} v
=
\partial_{y}(v^2)/2$,
equation \eqref{eq:MomentumBalanceFlow} becomes
\begin{equation}
\label{eq:MomentumBalanceFlow2}
\partial_{y} p_{\mathrm{aero}}
=
-
\frac{1}{2}\rho_a
\partial_{y}(v^2)
-
\frac{12\mu_a\,\mathrm{d}q_g}
{\mathrm{d}x\,d_g^3}.
\end{equation}
Let
 $v_{\mathrm{in}}(x,t)
=
{\mathrm{d}q_g(x,t)}/
(d_g(x,b,t)\,\mathrm{d}x)$ denote the entrance velocity at the inferior glottal boundary. Applying Bernoulli's equation between the subglottal reservoir and the glottal entrance yields $P_s
=
p_{\mathrm{aero}}(x,b,t)
+
\rho_a v_{\mathrm{in}}^2 /2$. Therefore,
\begin{equation}
\label{eq:EntrancePressure}
p_{\mathrm{aero}}(x,b,t)
=
P_s
-
\frac{1}{2}\rho_a
\left(
\frac{\mathrm{d}q_g}
{d_g(x,b,t)\,\mathrm{d}x}
\right)^2 .
\end{equation}
We now integrate \eqref{eq:MomentumBalanceFlow2} downstream from the inferior entrance $y=b$ to an arbitrary point $y\in[y_{\mathrm{sep}},b]$. Integrating both sides,
 reversing the integral limits, and rearranging yield
\begin{align}
p_{\mathrm{aero}}(x,y,t)
&=
p_{\mathrm{aero}}(x,b,t)
-
\frac{1}{2}\rho_a
\bigl(
v(x,y,t)^2-v_{\mathrm{in}}^2
\bigr)
\nonumber
\\[4pt]
&\quad
+
12\mu_a\,\mathrm{d}q_g
\int_y^b
\frac{1}
{\mathrm{d}x\,d_g(x,\eta,t)^3}
\,\mathrm{d}\eta .
\label{eq:ViscousBernoulliIntermediate}
\end{align}
Substituting \eqref{eq:EntrancePressure} together with
$v
=
{\mathrm{d}q_g}/
(d_g\,\mathrm{d}x)$,
gives
\begin{equation}
\label{eq:ViscousBernoulliPressure}
p_{\mathrm{aero}}(x,y,t)
=
P_s
-
\frac{1}{2}\rho_a
\left(
\frac{\mathrm{d}q_g}
{d_g(x,y,t)\,\mathrm{d}x}
\right)^2
+
12\mu_a\,\mathrm{d}q_g
\int_y^b
\frac{1}
{\mathrm{d}x\,d_g(x,\eta,t)^3}
\,\mathrm{d}\eta ,
\end{equation}
for $y_{\mathrm{sep}}
\le y\le b$. The upstream flow rate $\mathrm{d}q_g$ is determined by imposing the separation condition
$p_{\mathrm{aero}}(x,y_{\mathrm{sep}},t)=0$,
where the supraglottal pressure is assumed to vanish. Substituting $y=y_{\mathrm{sep}}$ into \eqref{eq:ViscousBernoulliPressure} gives
\begin{equation}
\label{eq:FlowRateQuadratic}
A_q(\mathrm{d}q_g)^2
-
B_q\mathrm{d}q_g
-
P_s
=
0,
\end{equation}
where
$$
A_q
=
\frac{1}{2}\rho_a
\frac{1}
{d_{\mathrm{sep}}^2(\mathrm{d}x)^2},~
B_q
=
12\mu_a
\int_{y_{\mathrm{sep}}}^{b}
\frac{1}
{\mathrm{d}x\,d_g(x,\eta,t)^3}
\,\mathrm{d}\eta .
$$
Therefore,
\begin{equation}
\label{eq:FlowRateViscousBernoulli}
\mathrm{d}q_g
=
\frac{
B_q
-
\sqrt{B_q^2+4A_qP_s}
}
{2A_q}.
\end{equation}
Once $\mathrm{d}q_g$ is computed, equation \eqref{eq:ViscousBernoulliPressure} determines the attached-flow pressure distribution for
$y_{\mathrm{sep}}
\le y\le b$. Downstream of separation, namely for $0\le y<y_{\mathrm{sep}},
$
the pressure is prescribed as  $p_{\mathrm{aero}}(x,y,t)=0$. 

If the glottis becomes locally closed, i.e., $d_{\min}=0,
$
we assume that no through-flow occurs. In this case,
\begin{equation}
\label{eq:AerodynamicPressureClosedRefined}
p_{\mathrm{aero}}
=
\begin{cases}
0,
&
0\le y<y_{\min},
\\[4pt]
P_s,
&
y_{\min}\le y\le b,
\end{cases}
\end{equation}
corresponding respectively to the downstream supraglottal side and the upstream subglottal side, and $\mathrm{d}q_g=0$. Finally, the total glottal upstream flow rate is obtained by integrating the slice flow rates along the longitudinal direction, i.e.,  $Q_g
=
-\int_0^L
\mathrm{d}q_g $.

\subsubsection*{Boundary conditions}

At the anterior ($x=0$) and posterior ($x=L$) margins, we impose kinematic boundary conditions corresponding to zero transverse displacement for both the membrane and beam components. Specifically, for all $t\geq0$ and all $y\in[0,b]$,
\begin{equation}
\label{eq:ZeroDisplacementMembraneBCs}
w_{\mathrm m}(0,y,t)=0,
\qquad
w_{\mathrm m}(L,y,t)=0,
\end{equation}
and, for the beam,
\begin{equation}
\label{eq:ZeroDisplacementBeamBCs}
w_{\mathrm b}(0,t)=0,
\qquad
w_{\mathrm b}(L,t)=0.
\end{equation}

For the beam component, we additionally prescribe bending-moment boundary conditions at the anterior and posterior margins. At the anterior margin ($x=0$), we assume that the surrounding tissue exerts a restoring moment proportional to the deviation of the local beam rotation from a prescribed rest angle $\theta_0$.

Under the zero shear strain assumption, the beam rotation is given by \eqref{eq:ZeroShearStrainBeam}.
Accordingly, the instantaneous angle between the medial plane and the beam centerline at $x=0$ is approximated by $\theta_G-\lambda^{3/2}\partial_x w_{\mathrm b}(0,t)$.  The anterior moment boundary condition is therefore given by
\begin{equation}
\label{eq:AnteriorMomentBC}
M_{c,\mathrm b}(0,t)
=
-
K_{r,a}
\left(
\theta_G
-
\lambda^{3/2}\partial_x w_{\mathrm b}(0,t)
-
\theta_0
\right),
\end{equation}
where $K_{r,a}>0$ denotes the anterior rotational stiffness coefficient.

Similarly, at the posterior margin ($x=L$), we assume a restoring moment proportional to the local beam rotation, yielding
\begin{equation}
\label{eq:PosteriorMomentBeamBC}
M_{c,\mathrm b}(L,t)
=
-
K_{r,p}
\lambda^{3/2}\partial_x w_{\mathrm b}(L,t),
\end{equation}
where $K_{r,p}\geq0$ denotes the posterior rotational stiffness coefficient.

Finally, we assume that the reference axis used in the definition of $M_{c,\mathrm b}$ coincides with the geometric center of the ligament layer, namely $r_{c,\mathrm b}
=
r_{\mathrm{lig}}$.

For the membrane component, we impose vanishing transverse shear forces at the inferior and superior boundaries ($y=0$ and $y=b$). This yields the homogeneous Neumann-type boundary conditions
\begin{equation}
\label{eq:ZeroShearMembraneBCs}
G_{\mathrm m}
\bigl(
\partial_y w_{\mathrm m}(x,\xi,t)
-
\varphi_{\mathrm m}(x,\xi,t)
\bigr)
+
\eta_{\mathrm m}
\bigl(
\partial_y\dot w_{\mathrm m}(x,\xi,t)
-
\dot\varphi_{\mathrm m}(x,\xi,t)
\bigr)
=
0,
\end{equation}
where $\xi\in\{0,b\}$,
$x\in[0,L]$,
 $t\geq0$. All the adopted numerical values of the stiffness and damping parameters of the  membrane-beam model are presented in Table \ref{tab:MechnicalParameters}.
 \begin{table}[h!]
\centering
\caption{\textbf{Stiffness and damping parameters. The shear stiffness and damping coefficients of the mucosa are known to vary considerably across experimental studies. The values adopted here are calibrated within the physiological ranges reported by \cite{ChanTitze99}.}}
\begin{tabular}{ccc}
\hline
Parameter & Value & Units \\
\hline
$K_{\mathrm{coup},\min}$ & 5.0$\times10^{4}$ & Pa/m \\
$K_{\mathrm{coup},\max}$ & 5.0$\times10^{5}$ & Pa/m \\
$C_{\mathrm{coup}}$ & 1.0$\times10^{1}$ & Pa.s/m \\
$K_{\mathrm{foun}}$ & 1.0$\times10^{2}$ & Pa \\
$C_{\mathrm{foun}}$ & 1.0 & Pa.s \\
$K_{\mathrm{col}}$ & 1.0$\times10^{6}$ & Pa \\
$C_{\mathrm{col}}$ & 1.0$\times10^{2}$ & Pa.s \\
$G_{\mathrm{m}}$ & 1.0$\times10^{2}$ & Pa \\
$\eta_{\mathrm{m}}$ & 1.0 & Pa.s \\
$K_{r,a}$ & 2.0$\times10^{-2}$ & N.m \\
$K_{r,p}$ & 1.0$\times10^{-2}$ & N.m \\
\hline
\end{tabular}
\label{tab:MechnicalParameters}
\end{table}

\subsubsection*{Acoustic output}

As illustrated in Fig.~\ref{fig:ModelOutline}, a one-way coupling is assumed between the VF dynamics and the acoustic field, whereby the radiated pressure is computed from the glottal flow rate. The acoustic field is simulated using the wave reflection analog (WRA) model \cite{KellyLochbaum62}. Consequently, the vocal fold model only provides the glottal flow input to the vocal tract.  Let $P_e$ denote the supraglottal pressure at the entrance of the vocal tract. This pressure is decomposed into the sum of outward- and inward-traveling components, $P_e = P_e^{+} + P_e^{-}$,
where $P_e^{+}$ represents the outward-traveling pressure wave and $P_e^{-}$ represents the incident (inward-traveling) pressure wave. At the entrance of the vocal tract, the wave variables satisfy
\[
P_e^{+}
=
P_e^{-}
+
\rho_a c_{a}\frac{Q_g}{A_e},
\]
where $c_{a}$ is the speed of sound and $A_e$ denotes the supraglottal cross-sectional area at the entrance of the vocal tract. The radiated pressure $p_{\mathrm{rad}}$ is obtained as an output of the WRA simulations.

\section*{Acknowledgments}
The authors used ChatGPT (OpenAI, GPT-5) to assist with language editing and 
clarity improvement. All technical content and results were verified by the authors. Research reported in this work was supported by the NIDCD of the NIH under Award No. P50DC015446, and ANID CIA250006. The content is solely the responsibility of the authors and does not necessarily represent the official views of the National Institutes of Health.
\section*{Author contributions}
\textbf{Conceptualization:} Mohamed A. Serry, Sean D. Peterson.
\paragraph*{Formal analysis:} Mohamed A. Serry.
\paragraph*{Supervision:} Mat\'ias Za\~nartu, Sean D. Peterson.
\paragraph*{Funding acquisition:} Mat\'ias Za\~nartu, Sean D. Peterson.

\paragraph*{Methodology:} Mohamed A. Serry.
\paragraph*{Software:} Mohamed A. Serry.
\paragraph*{Validation:} Mohamed A. Serry.
\paragraph*{Writing-original draft:} Mohamed A. Serry.
\paragraph*{Writing-review \& editing:} Mat\'ias Za\~nartu, Sean D. Peterson.\\


\begin{thebibliography}{10}

\bibitem{ChhetriNeubauerBerry12}
Chhetri DK, Neubauer J, Berry DA.
\newblock Neuromuscular control of fundamental frequency and glottal posture at phonation onset.
\newblock The Journal of the Acoustical Society of America. 2012;131(2):1401-12.

\bibitem{serry2023modeling}
Serry MA, Alzamendi GA, Za{\~n}artu M, Peterson SD.
\newblock Modeling the influence of the extrinsic musculature on phonation.
\newblock Biomechanics and modeling in mechanobiology. 2023;22(4):1365.

\bibitem{ZanartuGalindoErathPetersonWodickaHillman14}
Za{\~n}artu M, Galindo GE, Erath BD, Peterson SD, Wodicka GR, Hillman RE.
\newblock Modeling the effects of a posterior glottal opening on vocal fold dynamics with implications for vocal hyperfunction.
\newblock The Journal of the Acoustical Society of America. 2014;136(6):3262-71.

\bibitem{DejonckereKob09}
Dejonckere PH, Kob M.
\newblock Pathogenesis of vocal fold nodules: new insights from a modelling approach.
\newblock Folia Phoniatrica et Logopaedica. 2009;61(3):171-9.

\bibitem{HansonGerrattWard84}
Hanson DG, Gerratt BR, Ward PH.
\newblock Cinegraphic observations of laryngeal function in Parkinson's disease.
\newblock The Laryngoscope. 1984;94(3):348-53.

\bibitem{MorrisonRammage93}
Morrison MD, Rammage LA.
\newblock Muscle misuse voice disorders: description and classification.
\newblock Acta oto-laryngologica. 1993;113(3):428-34.

\bibitem{HillmanSteppVanStanZanartuMehta20}
Hillman RE, Stepp CE, Van~Stan JH, Za{\~n}artu M, Mehta DD.
\newblock An Updated Theoretical Framework for Vocal Hyperfunction.
\newblock American Journal of Speech-Language Pathology. 2020:1-7.

\bibitem{RajaeiBarzegarMojiriNilforoush14}
Rajaei A, Barzegar~Bafrooei E, Mojiri F, Nilforoush MH.
\newblock The occurrence of laryngeal penetration and aspiration in patients with glottal closure insufficiency.
\newblock International Scholarly Research Notices. 2014;2014.

\bibitem{SoderstenHertegardHammarberg95}
S{\"o}dersten M, Herteg{\aa}rd S, Hammarberg B.
\newblock Glottal closure, transglottal airflow, and voice quality in healthy middle-aged women.
\newblock Journal of Voice. 1995;9(2):182-97.

\bibitem{NguyenKennyTranLivesey09}
Nguyen DD, Kenny DT, Tran ND, Livesey JR.
\newblock Muscle tension dysphonia in Vietnamese female teachers.
\newblock Journal of Voice. 2009;23(2):195-208.

\bibitem{ChoiYeBerkeKreiman93}
Choi HS, Ye M, Berke GS, Kreiman J.
\newblock Function of the thyroarytenoid muscle in a canine laryngeal model.
\newblock Annals of Otology, Rhinology \& Laryngology. 1993;102(10):769-76.

\bibitem{ChhetriNeubauer15}
Chhetri DK, Neubauer J.
\newblock Differential roles for the thyroarytenoid and lateral cricoarytenoid muscles in phonation.
\newblock The Laryngoscope. 2015;125(12):2772-7.

\bibitem{YinZhang14}
Yin J, Zhang Z.
\newblock Interaction between the thyroarytenoid and lateral cricoarytenoid muscles in the control of vocal fold adduction and eigenfrequencies.
\newblock Journal of biomechanical engineering. 2014;136(11):111006.

\bibitem{YinZhang16}
Yin J, Zhang Z.
\newblock Laryngeal muscular control of vocal fold posturing: Numerical modeling and experimental validation.
\newblock The Journal of the Acoustical Society of America. 2016;140(3):EL280-4.

\bibitem{serry2023euler}
Serry MA, Alzamendi GA, Za{\~n}artu M, Peterson SD.
\newblock An Euler--Bernoulli-type beam model of the vocal folds for describing curved and incomplete glottal closure patterns.
\newblock Journal of the mechanical behavior of biomedical materials. 2023;147:106130.

\bibitem{MovahhediGengXueZheng21}
Movahhedi M, Geng B, Xue Q, Zheng X.
\newblock Effects of cricothyroid and thyroarytenoid interaction on voice control: Muscle activity, vocal fold biomechanics, flow, and acoustics.
\newblock The Journal of the Acoustical Society of America. 2021;150(1):29-42.

\bibitem{JiangGengZhengXue24}
Jiang W, Geng B, Zheng X, Xue Q.
\newblock A computational study of the influence of thyroarytenoid and cricothyroid muscle interaction on vocal fold dynamics in an MRI-based human laryngeal model.
\newblock Biomechanics and modeling in mechanobiology. 2024;23(5):1801-13.

\bibitem{TitzeStory02}
Titze IR, Story BH.
\newblock Rules for controlling low-dimensional vocal fold models with muscle activation.
\newblock The Journal of the Acoustical Society of America. 2002;112(3):1064-76.

\bibitem{AlzamendiPetersonErathHillmanZanartu21}
Alzamendi GA, Peterson SD, Erath BD, Hillman RE, Za{\~n}artu M.
\newblock Triangular body-cover model of the vocal folds with coordinated activation of the five intrinsic laryngeal muscles.
\newblock The Journal of the Acoustical Society of America. 2022;151(1):17-30.

\bibitem{TitzeHunter07}
Titze IR, Hunter EJ.
\newblock A two-dimensional biomechanical model of vocal fold posturing.
\newblock The Journal of the Acoustical Society of America. 2007;121(4):2254-60.

\bibitem{DeckerThomson07}
Decker GZ, Thomson SL.
\newblock Computational simulations of vocal fold vibration: Bernoulli versus Navier--Stokes.
\newblock Journal of Voice. 2007;21(3):273-84.

\bibitem{Zhang16}
Zhang Z.
\newblock Mechanics of human voice production and control.
\newblock The Journal of the Acoustical Society of America. 2016;140(4):2614-35.

\bibitem{alipour2011mathematical}
Alipour F, Brucker C, D~Cook D, Gommel A, Kaltenbacher M, Mattheus W, et~al.
\newblock Mathematical models and numerical schemes for the simulation of human phonation.
\newblock Current Bioinformatics. 2011;6(3):323-43.

\bibitem{Story08}
Story BH.
\newblock Comparison of magnetic resonance imaging-based vocal tract area functions obtained from the same speaker in 1994 and 2002.
\newblock The Journal of the Acoustical Society of America. 2008;123(1):327-35.

\bibitem{holmberg1989glottal}
Holmberg EB, Hillman RE, Perkell JS.
\newblock Glottal airflow and transglottal air pressure measurements for male and female speakers in low, normal, and high pitch.
\newblock Journal of Voice. 1989;3(4):294-305.

\bibitem{xue2012computational}
Xue Q, Mittal R, Zheng X, Bielamowicz S.
\newblock Computational modeling of phonatory dynamics in a tubular three-dimensional model of the human larynx.
\newblock The Journal of the Acoustical Society of America. 2012;132(3):1602-13.

\bibitem{chen2002electroglottographic}
Chen Y, Robb MP, Gilbert HR.
\newblock Electroglottographic evaluation of gender and vowel effects during modal and vocal fry phonation.
\newblock Journal of speech, language, and hearing research. 2002;45(5):821-9.

\bibitem{mehta2010voice}
Mehta DD, Deliyski DD, Zeitels SM, Quatieri TF, Hillman RE.
\newblock Voice production mechanisms following phonosurgical treatment of early glottic cancer.
\newblock Annals of Otology, Rhinology \& Laryngology. 2010;119(1):1-9.

\bibitem{StoryTitze95}
Story BH, Titze IR.
\newblock Voice simulation with a body-cover model of the vocal folds.
\newblock The Journal of the Acoustical Society of America. 1995;97(2):1249-60.

\bibitem{JiangLinHanson00}
Jiang J, Lin E, Hanson DG.
\newblock Vocal fold physiology.
\newblock Otolaryngologic Clinics of North America. 2000;33(4):699-718.

\bibitem{Titze88}
Titze IR.
\newblock The physics of small-amplitude oscillation of the vocal folds.
\newblock The Journal of the Acoustical Society of America. 1988;83(4):1536-52.

\bibitem{ThomsonMongeauFrankel05}
Thomson SL, Mongeau L, Frankel SH.
\newblock Aerodynamic transfer of energy to the vocal folds.
\newblock The Journal of the Acoustical Society of America. 2005;118(3):1689-700.

\bibitem{BhattacharyaSiegmund15}
Bhattacharya P, Siegmund T.
\newblock The role of glottal surface adhesion on vocal folds biomechanics.
\newblock Biomechanics and modeling in mechanobiology. 2015;14(2):283-95.

\bibitem{SmithTitze18}
Smith SL, Titze IR.
\newblock Vocal fold contact patterns based on normal modes of vibration.
\newblock Journal of biomechanics. 2018;73:177-84.

\bibitem{ChhetriNeubauerSoferBerry14}
Chhetri DK, Neubauer J, Sofer E, Berry DA.
\newblock Influence and interactions of laryngeal adductors and cricothyroid muscles on fundamental frequency and glottal posture control.
\newblock The Journal of the Acoustical Society of America. 2014;135(4):2052-64.

\bibitem{martinez2025toward}
Mart{\'\i}nez JD, Ibarra EJ, Parra JA, Mehta DD, Heaton JT, Hillman RE, et~al.
\newblock Toward Acoustic-Based Normalization of Laryngeal EMG for Improved Interspeaker Consistency in Muscle-to-Acoustic Mapping.
\newblock Journal of Voice. 2025.

\bibitem{AlzamendiManriquezHadwinDengPetersonErathMehtaHillmanZanartu20}
Alzamendi GA, Manr{\'\i}quez R, Hadwin PJ, Deng JJ, Peterson SD, Erath BD, et~al.
\newblock Bayesian estimation of vocal function measures using laryngeal high-speed videoendoscopy and glottal airflow estimates: An in vivo case study.
\newblock The Journal of the Acoustical Society of America. 2020;147(5):EL434-9.

\bibitem{TitzeAlipour06}
Titze I, Alipour F.
\newblock The Myoelastic-Aerodynamic Theory of Phonation.
\newblock National Center for Voice and Speech; 2006.

\bibitem{su2002measurement}
Su MC, Yeh TH, Tan CT, Lin CD, Linne OC, Lee SY.
\newblock Measurement of adult vocal fold length.
\newblock The Journal of Laryngology \& Otology. 2002;116(6):447-9.

\bibitem{lehoux2024methodology}
Lehoux S, Zhang Z.
\newblock A methodology to quantify the effective vertical thickness of prephonatory vocal fold medial surface.
\newblock Journal of Voice. 2024.

\bibitem{HunterTitze07}
Hunter EJ, Titze IR.
\newblock Refinements in modeling the passive properties of laryngeal soft tissue.
\newblock Journal of Applied Physiology. 2007;103(1):206-19.

\bibitem{chan1999viscoelastic}
Chan RW, Titze IR.
\newblock Viscoelastic shear properties of human vocal fold mucosa: Measurement methodology and empirical results.
\newblock The Journal of the Acoustical Society of America. 1999;106(4):2008-21.

\bibitem{JiangZhangStern01}
Jiang JJ, Zhang Y, Stern J.
\newblock Modeling of chaotic vibrations in symmetric vocal folds.
\newblock The Journal of the Acoustical Society of America. 2001;110(4):2120-8.

\bibitem{VanDenBergZantemaDoornenbal57}
Van~den Berg J, Zantema J, Doornenbal~Jr P.
\newblock On the air resistance and the Bernoulli effect of the human larynx.
\newblock The Journal of the Acoustical Society of America. 1957;29(5):626-31.

\bibitem{PelorsonHirschbergVanHasselWijnandsAuregan94}
Pelorson X, Hirschberg A, Van~Hassel R, Wijnands A, Auregan Y.
\newblock Theoretical and experimental study of quasisteady-flow separation within the glottis during phonation. Application to a modified two-mass model.
\newblock The Journal of the Acoustical Society of America. 1994;96(6):3416-31.

\bibitem{alipour2004flow}
Alipour F, Scherer RC.
\newblock Flow separation in a computational oscillating vocal fold model.
\newblock The Journal of the Acoustical Society of America. 2004;116(3):1710-9.

\bibitem{SidlofDoareCadotChaigne11}
{\v{S}}idlof P, Doar{\'e} O, Cadot O, Chaigne A.
\newblock Measurement of flow separation in a human vocal folds model.
\newblock Experiments in Fluids. 2011;51(1):123-36.

\bibitem{LuceroSchoentgen15}
Lucero JC, Schoentgen J.
\newblock Smoothness of an equation for the glottal flow rate versus the glottal area.
\newblock The Journal of the Acoustical Society of America. 2015;137(5):2970-3.

\bibitem{ChanTitze99}
Chan RW, Titze IR.
\newblock Viscoelastic shear properties of human vocal fold mucosa: Measurement methodology and empirical results.
\newblock The Journal of the Acoustical Society of America. 1999;106(4):2008-21.

\bibitem{KellyLochbaum62}
Kelly JL, Lochbaum CC.
\newblock Speech synthesis.
\newblock In: Proceedings of the Fourth International Congress on Acoustics; 1962. p. 127-30.

\end{thebibliography}

\end{document}